\pdfoutput=1

\documentclass[english,aps,pre,reprint]{revtex4-1}
\usepackage[T1]{fontenc}
\usepackage[latin9]{inputenc}
\usepackage{units}
\usepackage{amsmath}
\usepackage{amssymb}
\usepackage{graphicx}
\usepackage{epstopdf}

\usepackage{xcolor}
\usepackage{placeins}



\usepackage{babel}
\begin{document}

\title{Orientation of plastic rearrangements in two-dimensional model glasses under shear}

\author{Alexandre Nicolas} \email{alexandre.nicolas@polytechnique.edu}
\affiliation{ LPTMS, CNRS, Univ. Paris-Sud, Universit\'e Paris-Saclay,
  91405 Orsay, France }

\author{J\"org Rottler} \email{jrottler@physics.ubc.ca} \affiliation{
  Department of Physics and Astronomy and Quantum Matter Institute,
  University of British Columbia, Vancouver BC V6T 1Z1, Canada }

\begin{abstract}
The plastic deformation of amorphous solids is mediated by localized shear transformations involving 
small groups of particles rearranging irreversibly in an elastic background. We introduce and compare three different computational methods to extract the size and orientation of these shear transformations in simulations of a two-dimensional (2D) athermal model glass under simple shear. We find that the shear angles are broadly distributed around the macroscopic shear direction, with a more or less Gaussian distribution with a standard deviation of around $20^\circ$ about the direction of maximal local shear. The distributions of sizes and orientations of shear transformations
display no substantial sensitivity to the shear rate. These results can notably be used to refine the description of rearrangements in elastoplastic models.
\end{abstract}
\maketitle

\section{Introduction}
Polydisperse foams, highly concentrated emulsions, molecular glasses, and bulk metallic glasses exhibit microscopically heterogeneous mechanical properties.
As a result, these disordered solids do not deform
affinely under shear. Instead, their deformation features bursty rearrangements 
of small groups of particles embedded in an otherwise
elastically deforming medium. It is now well accepted that these microscopically localized shear transformations (ST)
are the elementary carriers of plastic deformation in sheared amorphous solids \cite{Argon1979,Falk1998}. 
By straining its surroundings, each ST
gives rise to a characteristic long-range deformation halo around it \cite{Maloney2006Amorphous,desmond2015measurement}, 
which mediates most collective effects in the material, such as cascades of rearrangements \cite{Baret2002,Antonaglia2014}.

Based on this picture at the particle scale, mesoscale elastoplastic 
models of amorphous plasticity have been
formulated, which divide the material into small regions (blocks) that 
are loaded elastically until they fail plastically \cite{nicolas2017deformation}. The failure
of a block is described as an ideal ST which partly dissipates the local stress and 
partly redistributes it to the other blocks. For an ST aligned with the principal direction of the macroscopic shear in  $d$-dimensional space, 
the Green's function $\mathcal{G}$ for the non-local redistribution
of the shear stress satisfies 
\begin{equation}
\label{eq:elastic_prop}
\mathcal{G}(r,\theta) \simeq C \cos [ 4\theta + 2\theta^\mathrm{pl} ] /r^{d}
\end{equation}
in the plane of the transformation, with a dimension-dependent prefactor $C$,
where $(r,\theta)$ are the polar coordinates in the frame centered on the plastic block and $\theta^\mathrm{pl}$ (defined precisely in Eq.~\eqref{eq:eigenstrain_Esh}) refers to
the orientation of the individual ST. The far field limit of this expression for $\mathcal{G}$ matches Eshelby's solution
for a spherical inclusion endowed with a spontaneous strain \cite{Eshelby1957}, and was shown to suitably describe
the disorder-averaged response of an amorphous solid to an ideal ST in atomistic simulations \cite{Puosi2014}.

Mesoscale models, however, rest on several assumptions concerning the STs, including their idealized "Eshelby" nature, their equal size, and their orientation along the direction of maximal local shear \cite{Nicolas2014u,sandfeld2014deformation}, 
or even along the macroscopic shear direction in scalar models \cite{Talamali2011,Budrikis2013} (in this regard, ref.~\cite{Homer2009} is an exception).
To give them stronger footing,
experimental and numerical efforts have been made to characterize plastic rearrangements,  as exposed in Sec.~\ref{sec:state_of_the_art}. In particular, 
much attention has been paid to their shape and their size \cite{Argon1979,Schall2007,ma2015nanoindentation,albaret2016mapping},
while the question of their orientation has remained largely unexplored, despite its obvious  relevance for the buildup of spatial correlations between individual STs \cite{Nicolas2014s,lemaitre2015tensorial}.
In this contribution, we simulate the shear deformation of a two-dimensional (2D) athermal model glass (described in
Sec.~\ref{sec:model}) with molecular dynamics in order to study the
statistical properties of actual rearrangements for different shear rates. Strong emphasis is placed on their angles of failure.
To this end, we propose (in Sec.~\ref{sec:model}) and compare (in Sec.~\ref{sec:comparison_methods}) several numerical methods to extract these angles.
We find that these angles are broadly distributed around the macroscopic shear direction, with a more or less Gaussian distribution
with a standard deviation of around $20^\circ$. Overall, the sizes and orientations of the detected rearrangements are 
fairly insensitive to the shear rate, but many of them actually differ from ideal STs. Even when the ideal ST description works reasonably well,
local methods relying exclusively on the displacements (or forces) of the most active rearranging particles  give poor estimates of the ST orientation; the latter is recovered if a broader selection of particles near the ST is considered. 

\section{Previous endeavors to characterize plastic rearrangements \label{sec:state_of_the_art} }
Leaving aside Schwarz's early attempts to classify rearrangements in a 3D foam at rest \cite{schwarz1965rearrangements}, Argon and Kuo
were the first to report localized rearrangements in a disordered system, more precisely a 2D foam (`bubble raft') that was used as a model system
for metallic glasses \cite{Argon1979}. Interestingly, they mentioned two types of STs: sharp slips of rows of about 5 bubbles in length and more diffuse cooperative
rearrangements of regions of 5 bubbles in diameter. In the 1980's, Princen studied the swap of neighbors between four bubbles (in 2D) to
account for some rheological properties of foams and concentrated emulsions \cite{Princen1983}; the detailed dynamics of this swap process
were investigated much later in clusters of 4 bubbles \cite{biance2011strain}. In slowly sheared colloidal glasses, STs were directly visualized 
using confocal microscopy and their core was observed to be around 3 particle diameters in linear size \cite{Schall2007}. In metallic glasses, direct visualization of
STs cannot be achieved experimentally but estimates for their volumes can be obtained indirectly (e.g., via nano-indentation tests and their
sensitivity to the shear rate) and typically
correspond to a few dozen atoms ($\sim 30$ in the Zr-based glass studied with nano-indentation tests in \cite{choi2012indentation}), with 
a possible dependence on the sample morphology (for instance, for a Ni-Nb metallic glass, the ST size was reported to decrease from 83 atoms
to 36 atoms when the material was cast into a $\mu$m-thin film \cite{ma2015nanoindentation}). 

Numerically, the most comprehensive characterization of rearrangements to date was performed by Albaret et al. \cite{albaret2016mapping}
on a 3D atomistic model for amorphous bulk silicon under quasi-static shear. Rearrangements were detected by artificially
reverting the applied strain increments at every step and deducing the irreversible changes that took place; the detected rearrangements
were then modeled as a collection of Eshelby inclusions, whose sizes (or volumes $V_0$) and eigenstrains $\boldsymbol{\epsilon}^\star$ were fitted 
to best reproduce the displacement field measured during the actual strain increment. These inclusions were shown to account for all
plastic effects visible in the stress--strain curves of these materials and the effective volume $\gamma^\star V_0$ (where $\gamma^\star$
is the
maximal shear component of $\boldsymbol{\epsilon}^\star$) was found to be exponentially distributed, with a typical size of $70\, \mathring{\mathrm{A}}^3$, while both dilational and contractional volumetric strains were observed. 
The evolution of the effective volume  $\gamma^\star V_0$ during the transformation was computed in \cite{boioli2017shear} by detecting the saddle point; the value of the effective volume at this saddle point, called
activation volume, was found to amount to around $20\%$ of the final $\gamma^\star V_0$.

\section{Numerical model and methods \label{sec:model}}
\begin{figure}
\includegraphics[width=0.9\linewidth]{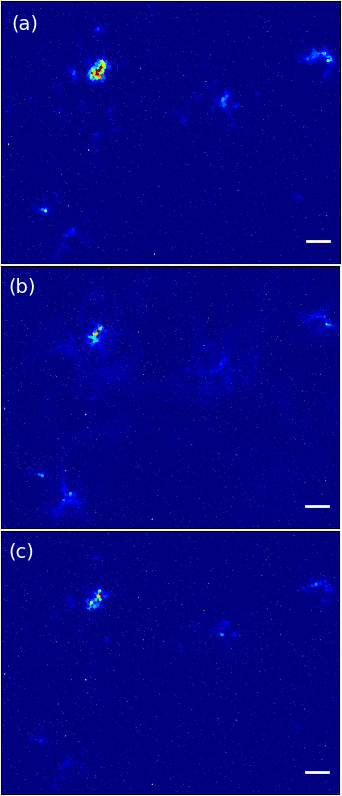}
\caption{\label{event-fig}Detection of plastic events via \textbf{(a)} the $d^2_{\rm min}$ criterion, \textbf{(b)} by the average kinetic energy of a particle and \textbf{(c)}  the magnitude of the linearized "Hessian" forces (see text). The scale bar is 10 particle diameter.}
\end{figure}

\subsection {Model and simulation protocol}

In order to get information on the morphology and orientation of STs,
we perfom molecular dynamics simulations of an amorphous material (a glass) under simple shear,
in 2D and in the athermal limit. The model glass is a binary mixture of A and B particles, with
$N_{A}=32500$ and $N_{B}=17500$, of respective diameters
$\sigma_{AA}=1.0$ and $\sigma_{BB}=0.88$, confined in a square box of
dimensions $205\sigma_{AA}\times205\sigma_{AA}$, with periodic
boundary conditions. The system,  at density 1.2, 
was prepared by quenching an equilibrated configuration at temperature
$T= 1$ with a fast quenching rate $\frac{dT}{dt} = 2 \cdot 10^{-3}$, at constant volume. The
particles, of mass $m=1$, interact via a pairwise Lennard-Jones
potential,
\[
V_{\alpha\beta}\left(r\right)=4\epsilon_{\alpha\beta}\left[\left(\frac{\sigma_{\alpha\beta}}{r}\right)^{12}-\left(\frac{\sigma_{\alpha\beta}}{r}\right)^{6}\right],
\]
where $\alpha,\beta=A,\,B$, $\sigma_{AB}=0.8$,$\epsilon_{AA}=1.0$,
$\epsilon_{AB}=1.5$, and $\epsilon_{BB}=0.5$. The potential is truncated
at $r=2.5\sigma_{AA}$ and shifted for continuity. Simple shear $\gamma$
is imposed at rate $\dot{\gamma}$ by deforming the (initially square) box into
a parallelogram
and remapping the particle positions. After an initial transient (20\% strain), 
the system reaches a steady state, which is the focus of the present study.

In the athermal limit, the equations of motion read
\begin{equation}
\frac{d r_{i}}{dt}  =  v_{i}\nonumber;\ \ \ 
m\frac{d v_{i}}{dt}  =  -\sum_{i\neq j}\frac{\partial V\left(r_{ij}\right)}{\partial r_{ij}}+f_{i}^{D}.\label{eq:eq_of_motion_MD}
\end{equation}
The dissipative force $f_{i}^{D}$
experienced by particle \emph{i} is computed with a Dissipative Particle
Dynamics scheme, viz.,
\begin{eqnarray}
f_{i}^{D} & = & -\sum_{j\neq i}\zeta w^{2}\left(r_{ij}\right)\frac{v_{ij}\cdot r_{ij}}{r_{ij}^{2}} r_{ij}\label{eq:f_diss_DPD}\\
\text{where }w(r) & \equiv & \begin{cases}
1-\frac{r}{r_{c}} & \text{ if }r<r_{c} \equiv 3\sigma_{AA},\\
0 & \text{ otherwise.}
\end{cases}\nonumber 
\end{eqnarray}
Here, $v_{ij}\equiv v_{i}- v_{j}$
denotes the relative velocity of particle $i$ with respect to $j$,
$r_{ij}\equiv r_{i}- r_{j}$, and $\zeta=1/\tau_{LJ}$ controls
the damping intensity (the effect of the damping was studied in \cite{nicolas2016effects}). Equations \eqref{eq:eq_of_motion_MD} are integrated
with the velocity Verlet algorithm with a time step $dt=0.005$. In all the
following, we use $\tau_{LJ}\equiv\sqrt{m\sigma_{AA}^{2}/\epsilon}$ as
the unit of time and $\sigma_{AA}$ as the unit of length.

\begin{figure}[t]
\noindent \begin{centering}
\includegraphics[width=0.4\columnwidth]{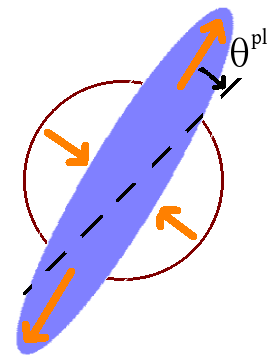} 
\par\end{centering}
\caption{\label{fig:sketch_angle}Representation of the angle of failure $\theta^\mathrm{pl}$. The orange arrows indicate the elongational
and contractional directions of an ideal ST, while the dashed line represents the elongational direction of the macroscopic shear.  }
\end{figure}

\begin{figure}
\begin{centering}
\includegraphics[width=\linewidth]{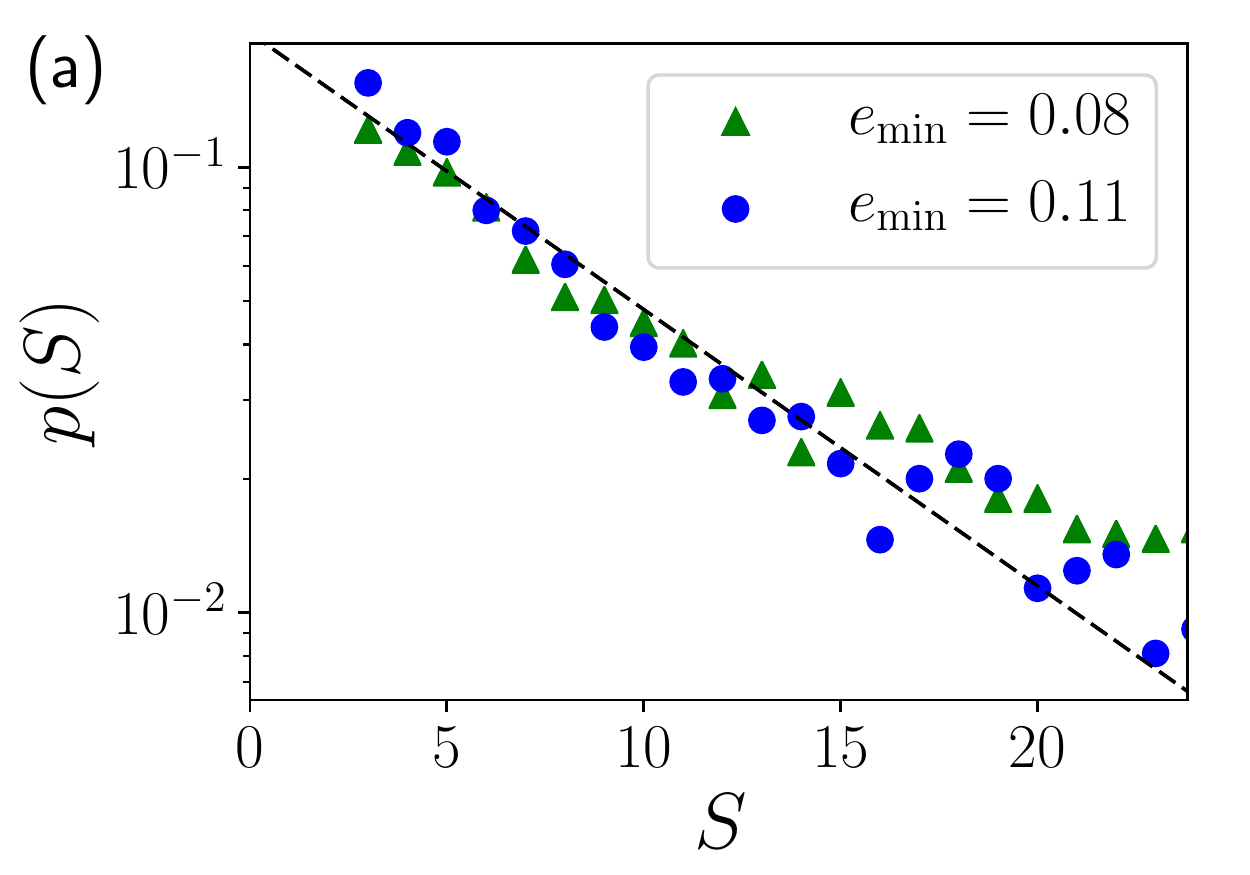}
\includegraphics[width=\linewidth]{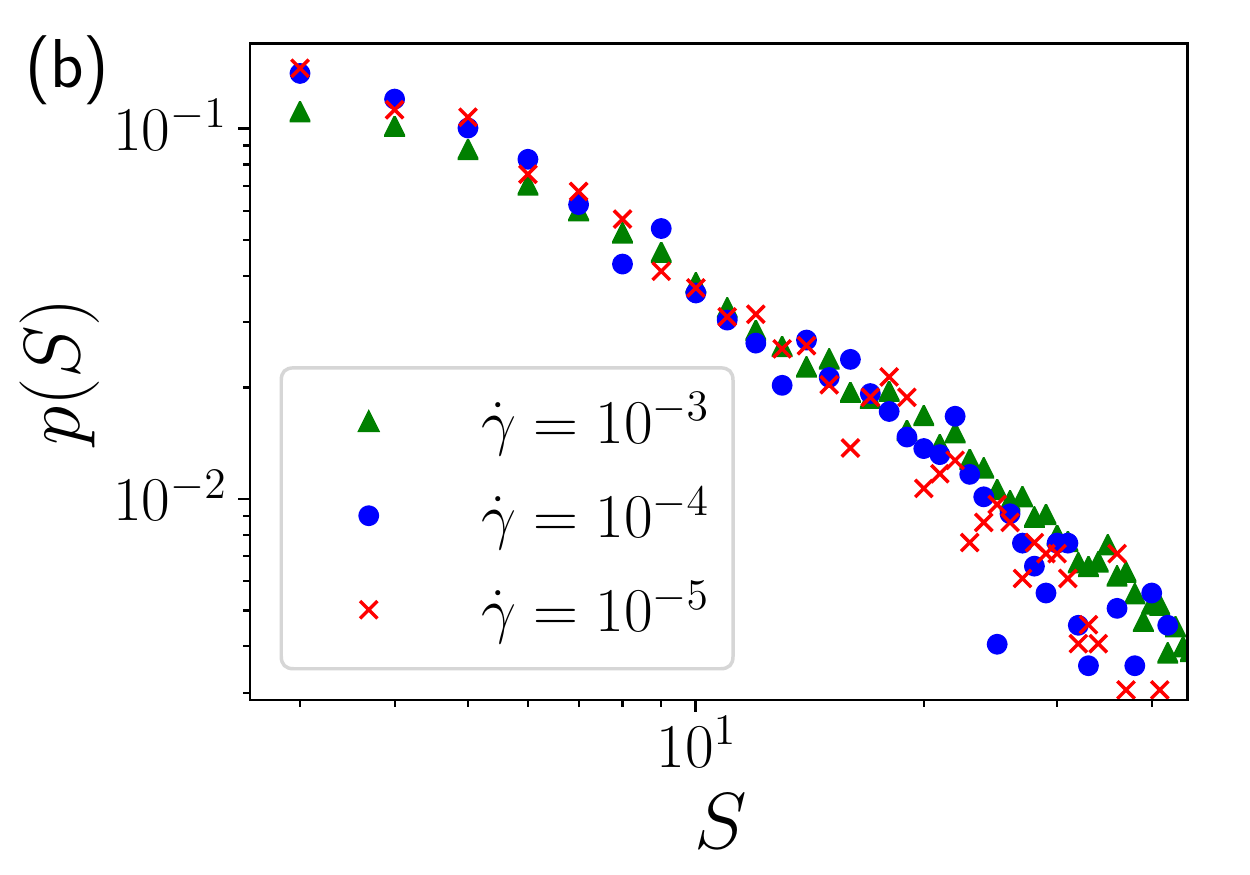}
\par\end{centering}
\caption{\label{fig:size_vs_threshold} Distribution of sizes $S$ of the rearranging clusters detected with the kinetic energy based criterion \textbf{(a)} for two different threshold values $e_\mathrm{min}$ at $\dot\gamma=10^{-5}$ and \textbf{(b)}
for three different shear rates $\dot\gamma$ with $e_\mathrm{min}=0.11$. The thin dashed line in the top panel is proportional to $\mathrm{exp}(-S/S_0)$ with $S_0=7$.}
\end{figure}

\subsection{Detection of rearrangements \label{sec:detection_rearr}}

As expected, the simulations display fast localized rearrangements. 
Several measures are available to identify them and are known to yield comparable 
results \cite{Chikkadi2012b}. In Fig.~\ref{event-fig}, we compute three of these diagnostics
of non-affinity on a typical snapshot of a simulation at shear rate
$\dot{\gamma}=10^{-4}$. These diagnostics are based on the displacements $\delta u_j$ of particles $j$ during a short time interval 
$[t,\,t+\delta t]$, with
$\delta t=2$. Panel (a) shows the amplitude of the minimized
mean-square difference 
\begin{equation*}
d^2_{\rm min} = \min_{\boldsymbol{G}} \sum_{r_j \in \mathcal{C}} \left[\delta u_j - \delta u_0 - \boldsymbol{G}\cdot (r_j - r_0) \right]^2
\end{equation*}
between the actual displacements $\delta u_j$ of particles $j$ in a circular region $\mathcal{C}$ around a
given particle $r_0$ and \emph{any} set of affine displacements, i.e., displacements
resulting from a uniform displacement gradient $\boldsymbol{G}$ during $\delta t$ \cite{Falk1998}. This measure of the nonaffine residual
strain has become a quasi gold standard for identifying plastic
rearrangements in amorphous solids. Panel (b) shows a simpler
measure, namely the amplitude of the average kinetic energy of a
particle averaged over $\delta t$. The motivation is that in an
athermal system, only particles undergoing a rearrangement are expected to have large marginal
velocities. Lastly, in panel (c) we consider 
the magnitudes of the (linearized) forces $ f^{(\mathcal{H})}_i=-\sum_j {\boldsymbol{\mathcal{H}}_{ij}(t)} \cdot {\delta u_j}$,
where $\boldsymbol{\mathcal{H}}_{ij}(t)= \frac{\partial^2 V}{\partial r_i r_j}$ is the Hessian matrix at time $t$. These 
are the forces that effectively drive plastic rearrangements. As discussed by Lem{a\^i}tre \cite{lemaitre2015tensorial}, 
they also localize in regions of high non-affine strain.

Figure~\ref{event-fig} confirms that the three methods studied give very similar results. Accordingly,
for convenience, we choose to use a criterion based on kinetic energies to detect rearrangements. More precisely,
particles with a kinetic energy larger than an arbitrary threshold $e_\mathrm{min}$ are considered to be rearranging; the threshold
value is lowered to $\nicefrac{3}{4}\, e_\mathrm{min}$ for the neighbors of rearranging particles, in order to obtain more compact
ST shapes, where two particles are defined as neighbors if they are separated by a distance smaller than 2. Finally, rearranging
particles are partitioned into clusters of neighbors, each corresponding to an individual ST (clusters with fewer than 3 particles were discarded). The distributions $p(S)$ of sizes of the resulting clusters for distinct thresholds
$e_\mathrm{min}$ and distinct shear rates $\dot\gamma$ are represented on Fig.~\ref{fig:size_vs_threshold}; neither the threshold nor
the shear rate seem to considerably alter the seemingly slower-than-exponential (but faster-than-power-law) decay of $p(S)$.
In the following, we shall see that all our results are fairly insensitive to these parameters $e_\mathrm{min}$ and $\dot\gamma$.
We have also checked (though inexhaustively) that the distributions of orientations of rearrangements detected on the basis of the linearized forces $ f^{(\mathcal{H})}_i$ are compatible with those shown below.

\subsection{Methods to measure ST orientations}
\begin{figure*}[p]
\noindent \begin{centering}
\includegraphics[width=0.37\textwidth]{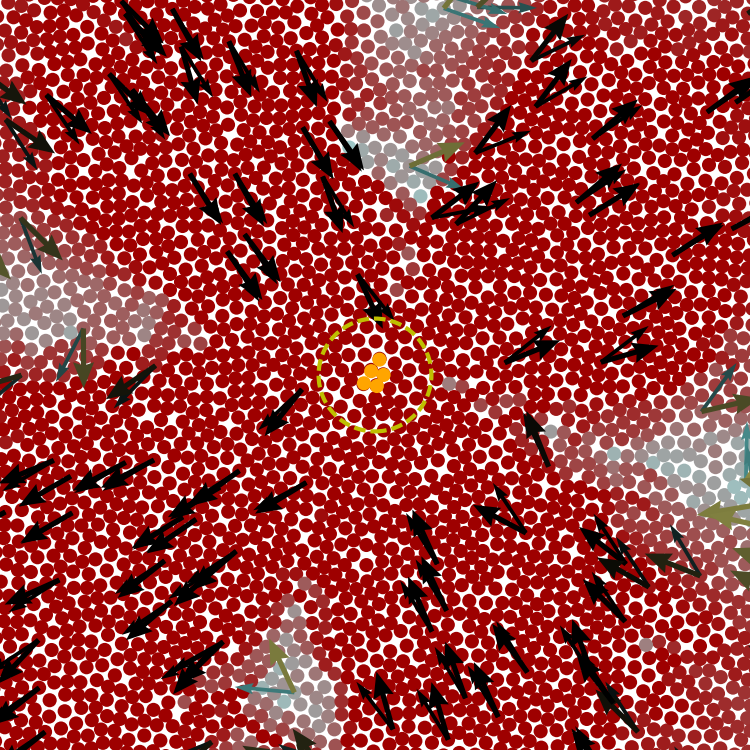} \hspace{0.2cm} \includegraphics[width=0.45\textwidth]{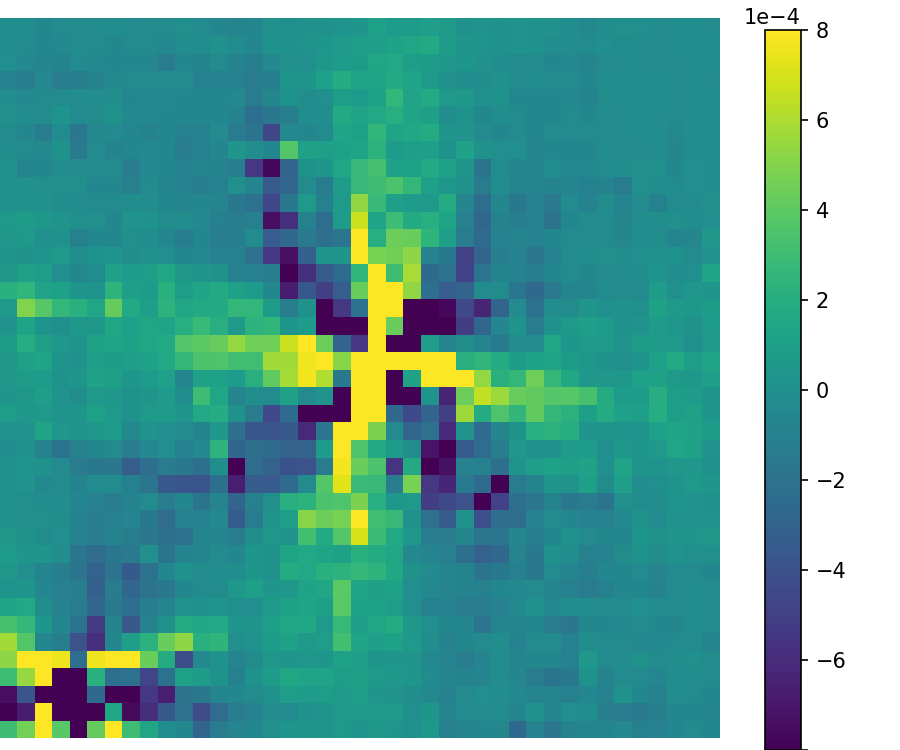}
\vspace{0.3cm}

\includegraphics[width=0.37\textwidth]{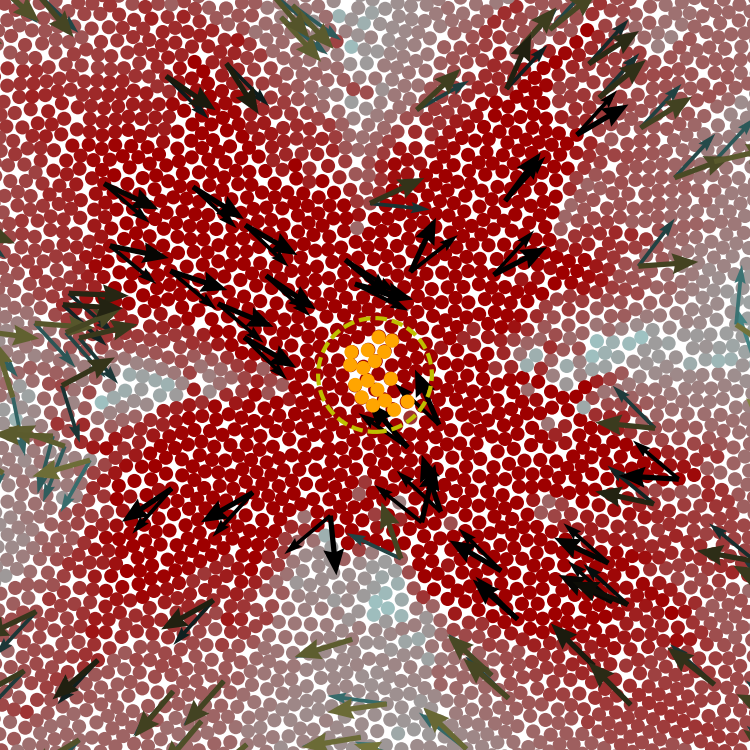}  \hspace{0.2cm} \includegraphics[width=0.45\textwidth]{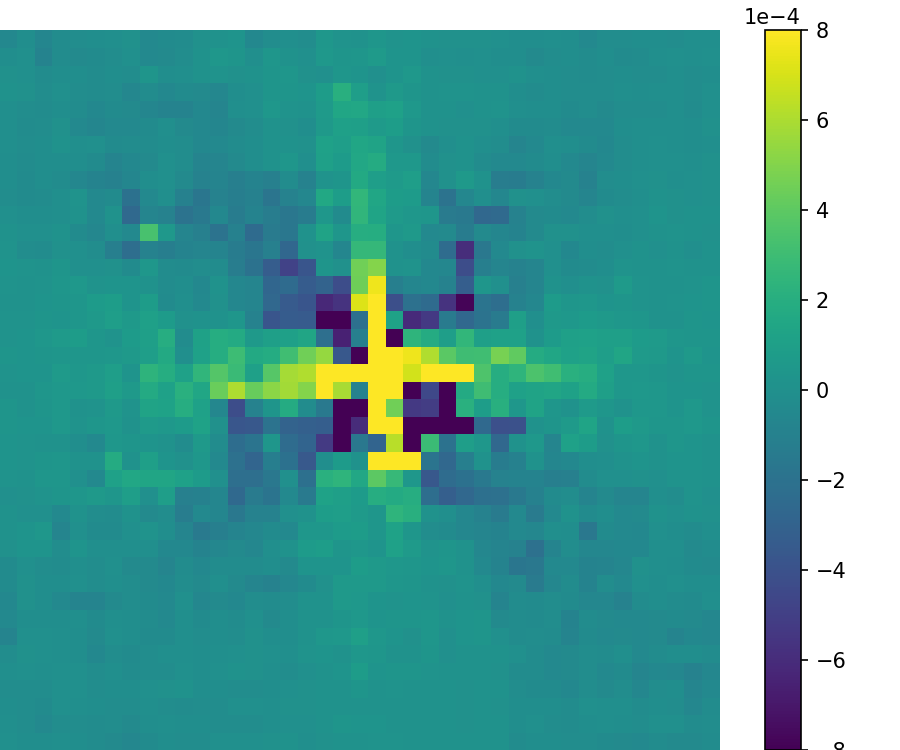}
\vspace{0.3cm}

\includegraphics[width=0.37\textwidth]{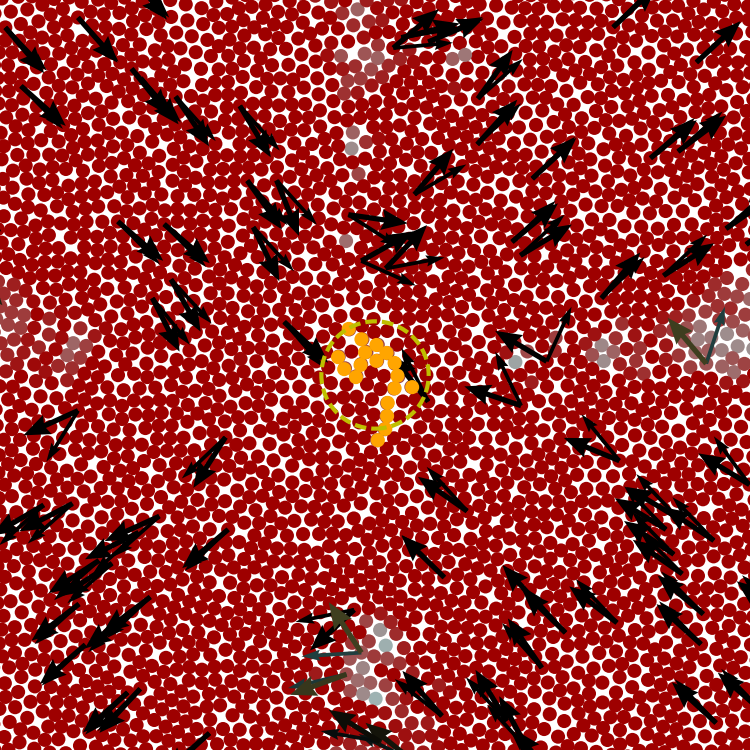}  \hspace{0.2cm} \includegraphics[width=0.45\textwidth]{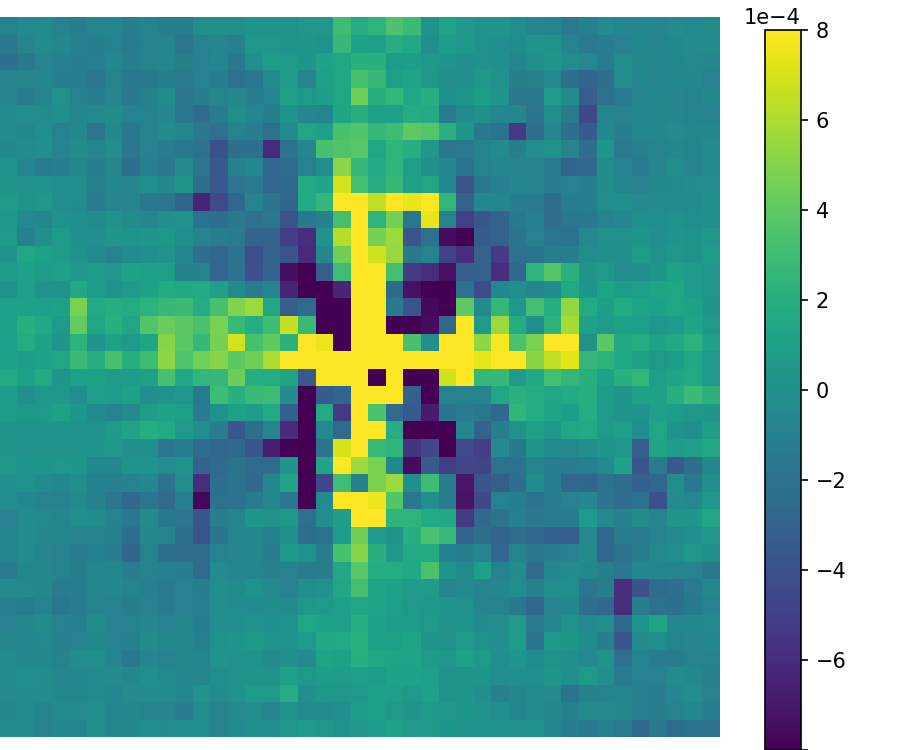}
\par\end{centering}

\caption{ \label{fig:snapshots} Elastic reponse computed in the auxiliary MD simulations (see text) to a selection of three STs exhibiting a quadrupolar response.  
In the left column, particles in the ST are colored in orange, while the colors of the other particles depend on the norms of their
displacements $\delta u_i$ (warmer colors denote larger displacements). The arrows with wide shafts represent the directions of
 $\delta u_i$ for a random subset of particles, while the (directions of) displacements  represented by narrower arrows are the response to the 
 best-fitting Eshelby inclusion. The figures shown are zooms on a $50\times 50$ portion of the global system (of size $205 \times 205$). The right column presents the coarse-grained strain field $\delta \epsilon_{xy}^c$ computed from the associated auxiliary simulations, in a $100\times 100$ square around the cluster. }
\end{figure*}

\begin{figure}[t]
\noindent \begin{centering}
\includegraphics[width=\columnwidth]{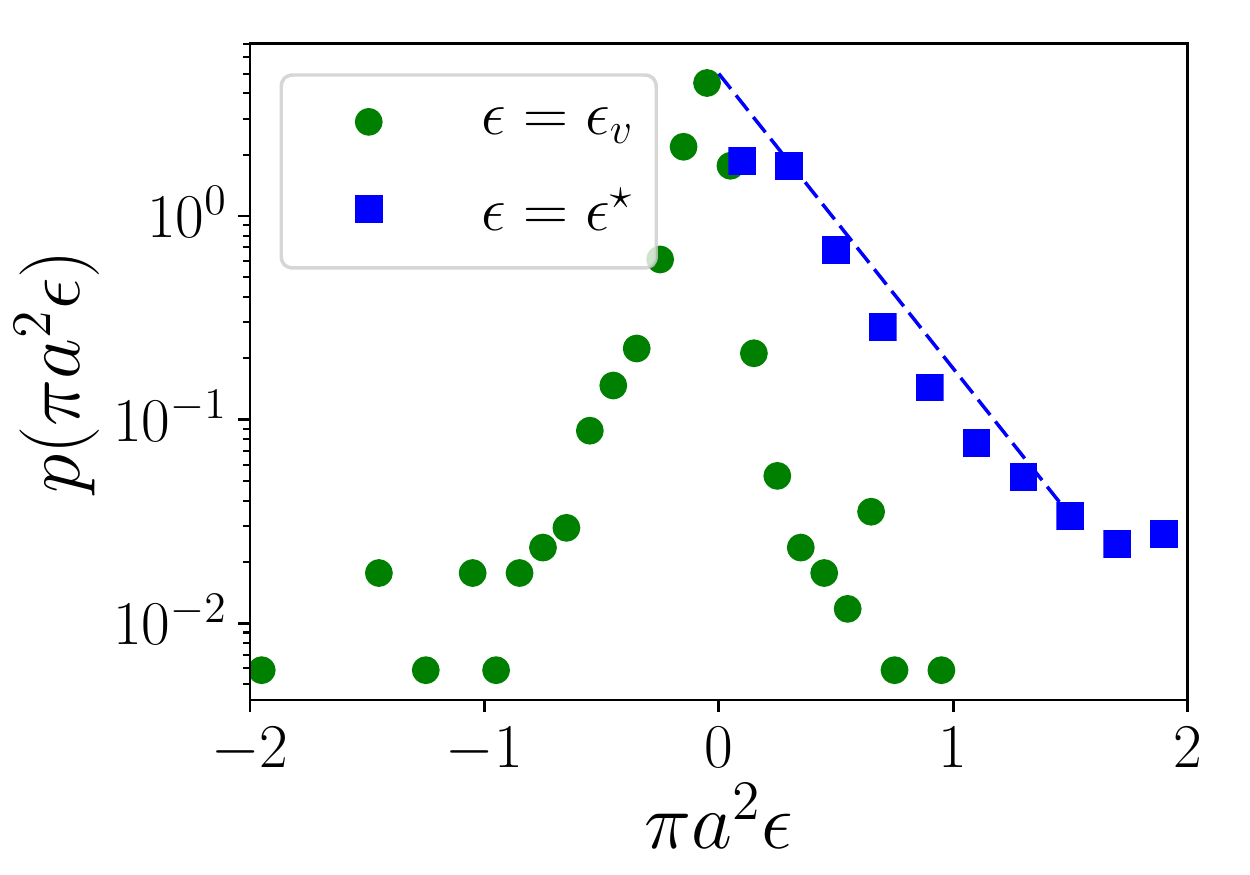} 
\par\end{centering}
\caption{\label{fig:dist_strain_strenghts}Distribution of the dilational strengths $\pi a^2 \epsilon_v$ (circles) and the shear strengths $\pi a^2 \epsilon^\star$ (squares)
of the STs detected at $\dot\gamma=10^{-5}$ (with threshold $e_\mathrm{min}=0.11$). The dashed blue line is proportional to $\exp(-x/0.3)$}
\end{figure}

In order to study ST orientations, 
a rearrangement is likened to a circular Eshelby inclusion with an eigenstrain $\boldsymbol{\epsilon}^\star$,
 i.e., a region whose stress-free state is not reached for a deformation
$\boldsymbol{\epsilon}(r)=\boldsymbol{0}$, but for $\boldsymbol{\epsilon}(r)=\boldsymbol{\epsilon}^\star$ (if it were unconstrained).
The eigenstrain $\boldsymbol{\epsilon}^\star$ can be split into a deviatoric part, associated with shape change, and a volumetric part, associated
with local dilation, \emph{viz.},
\begin{equation}
\label{eq:eigenstrain_Esh}
\boldsymbol{\epsilon}^\star=
\epsilon^{\star}\left(\begin{array}{cc}
\sin 2\theta^{\mathrm{pl}} & \cos 2\theta^{\mathrm{pl}} \\
 \cos 2\theta^{\mathrm{pl}}  & -\sin 2\theta^{\mathrm{pl}}
\end{array}\right)+\epsilon_{v}\left(\begin{array}{cc}
1 & 0\\
0 & 1
\end{array}\right)
\end{equation}
with $\epsilon^\star \geqslant 0$. We define the ST orientation as the \emph{angle of failure} $\theta^{\mathrm{pl}} \in ]-90^\circ,90^\circ]$; it is thus the angle between
the elongational principal direction of the ST and that of the macroscopic shear, as sketched in Fig.~\ref{fig:sketch_angle}.

\subsubsection{Fit to an Eshelby inclusion}

We are now left with the problem of determining $\boldsymbol{\epsilon}^\star$ in practice. Drawing 
inspiration from Albaret et al. \cite{albaret2016mapping}, we exploit the elastic field induced
by an inclusion \emph{\`a la} Eshelby. For homogeneous isotropic elastic media, the deformation $\boldsymbol{\epsilon}^\mathrm{in}$
within any embedded elliptical inclusion will be constant. It naturally follows that, for a circular inclusion, the principal
directions of $\boldsymbol{\epsilon}^\mathrm{in}$ and $\boldsymbol{\epsilon}^\star$ will be identical, owing to symmetry arguments.
Outside the circular inclusion (of radius $a$ and centered at $r=0$), the induced displacements $\delta u$ are given by \cite{jin2017displacement}
\begin{widetext}
\begin{eqnarray}
\label{eq:Eshelby_displacements}
\delta u_{1}(r) & = & \frac{x_{1}}{8(1-\nu)}\tilde{a}^{2}\left\{ \left[2(1-2\nu)+\tilde{a}^{2}\right]\left(\epsilon_{11}-\epsilon_{22}\right)+2\tilde{a}^{2}\left(\epsilon_{11}+\epsilon_{22}\right)+4\left(1-\tilde{a}^{2}\right)\left(\tilde{x}_{1}^{2}\epsilon_{11}+\tilde{x}_{2}^{2}\epsilon_{22}\right)\right\} \\ \nonumber
 &  & +\frac{x_{2}}{8(1-\nu)}\tilde{a}^{2}\cdot2\epsilon_{12}\left[2(1-2\nu)+\tilde{a}^{2}+4\left(1-\tilde{a}^{2}\right)\tilde{x}_{1}^{2}\right]
 \\ \nonumber
\delta u_{2}(r) & = & \frac{x_{2}}{8(1-\nu)}\tilde{a}^{2}\left\{ \left[2(1-2\nu)+\tilde{a}^{2}\right]\left(\epsilon_{22}-\epsilon_{11}\right)+2\tilde{a}^{2}\left(\epsilon_{11}+\epsilon_{22}\right)+4\left(1-\tilde{a}^{2}\right)\left(\tilde{x}_{1}^{2}\epsilon_{11}+\tilde{x}_{2}^{2}\epsilon_{22}\right)\right\} \\ \nonumber
 &  & +\frac{x_{1}}{8(1-\nu)}\tilde{a}^{2}\cdot2\epsilon_{12}\left[2(1-2\nu)+\tilde{a}^{2}+4\left(1-\tilde{a}^{2}\right)\tilde{x}_{2}^{2}\right],
\end{eqnarray}

\end{widetext}
where $r=(x_1,x_2)$ and tildes denote distances rescaled by the norm of $r$ (viz., $\tilde x_1= x_1/r$).

For each rearranging cluster, the equivalent size $a$ and eigenstrain components $\epsilon^\star$, $\epsilon_v$, and $\theta^\mathrm{pl}$
defined in Eq.~\eqref{eq:eigenstrain_Esh} are calculated as the parameters minimizing
the squared difference between the particle displacements $\delta u^\prime_i$ over $\delta t=2$ and the theoretical
expectations of Eq.~\eqref{eq:Eshelby_displacements}, for all particles $i$ that are at a distance between $2a$ and a large distance $d_\mathrm{max}$ 
away from the cluster center; the quality of the fit will be measured by the relative squared difference $\chi^2$. (Note that the results turned out to be insensitive to the value of $d_\mathrm{max}$.) However, unlike ref.~\cite{albaret2016mapping},
the displacements $\delta u^\prime_i$ are not extracted from the actual dynamical simulation. Instead, in order to avoid the superposition of many STs, we run an auxiliary
simulation for each rearranging cluster so as to measure the response induced only by this cluster. Pragmatically, starting from the configuration
 at $t$, we move particles $j$ belonging to the cluster by a fraction $\alpha \ll 1$ of their actual displacements $\delta u_j$, pin them to their
 new positions and obtain the response $\alpha \delta u^\prime_i$ of the other particles to this local rearrangement by minimization This strategy, which we refer to as MD/Esh, 
 will be our main method to access the ST morphology. 
One should nevertheless be aware that the results of the auxiliary simulations display
a slight sensitivity to the details of the minimization procedure, but the
consistency of our results will prove that this sensitivity can be overlooked.

\begin{figure}[t]
\noindent \begin{centering}
\includegraphics[width=\columnwidth]{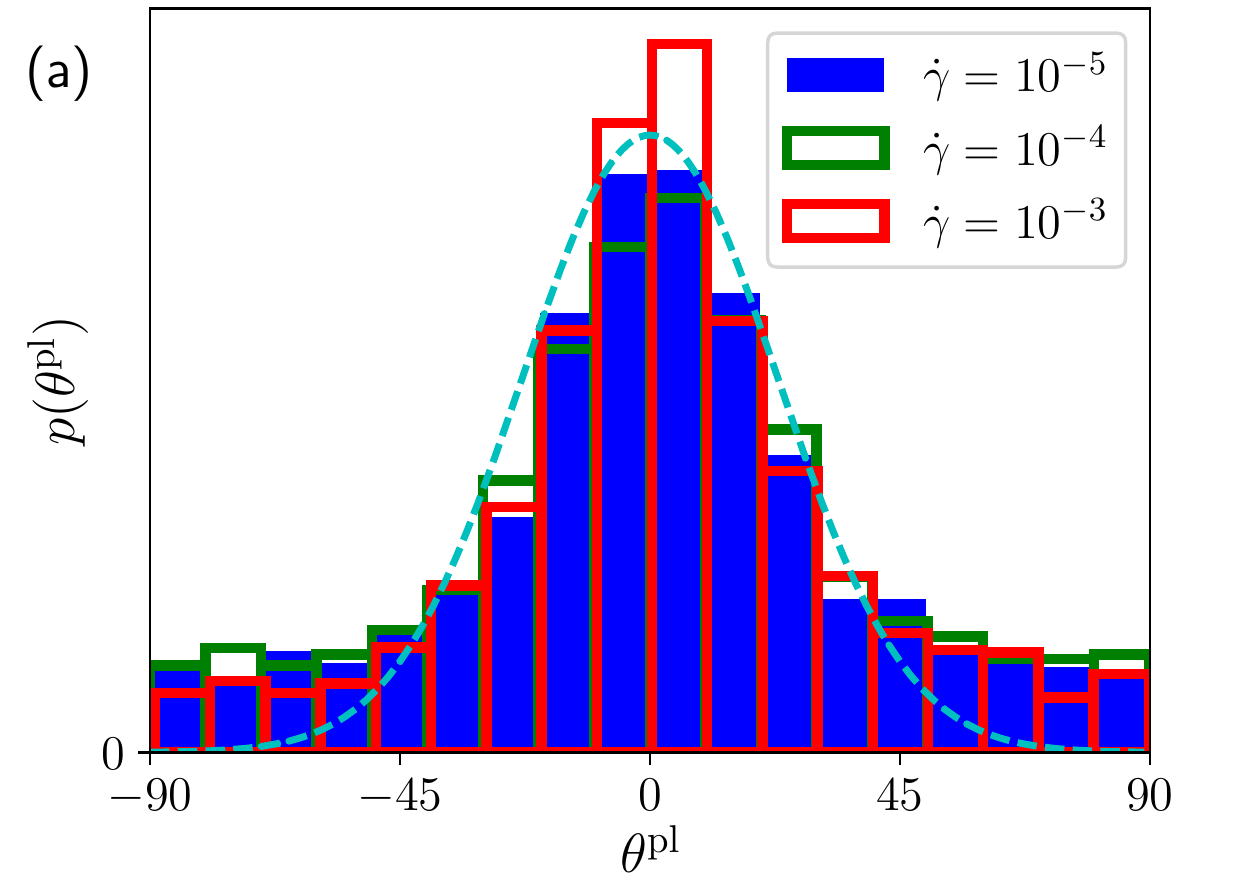}
\includegraphics[width=\columnwidth]{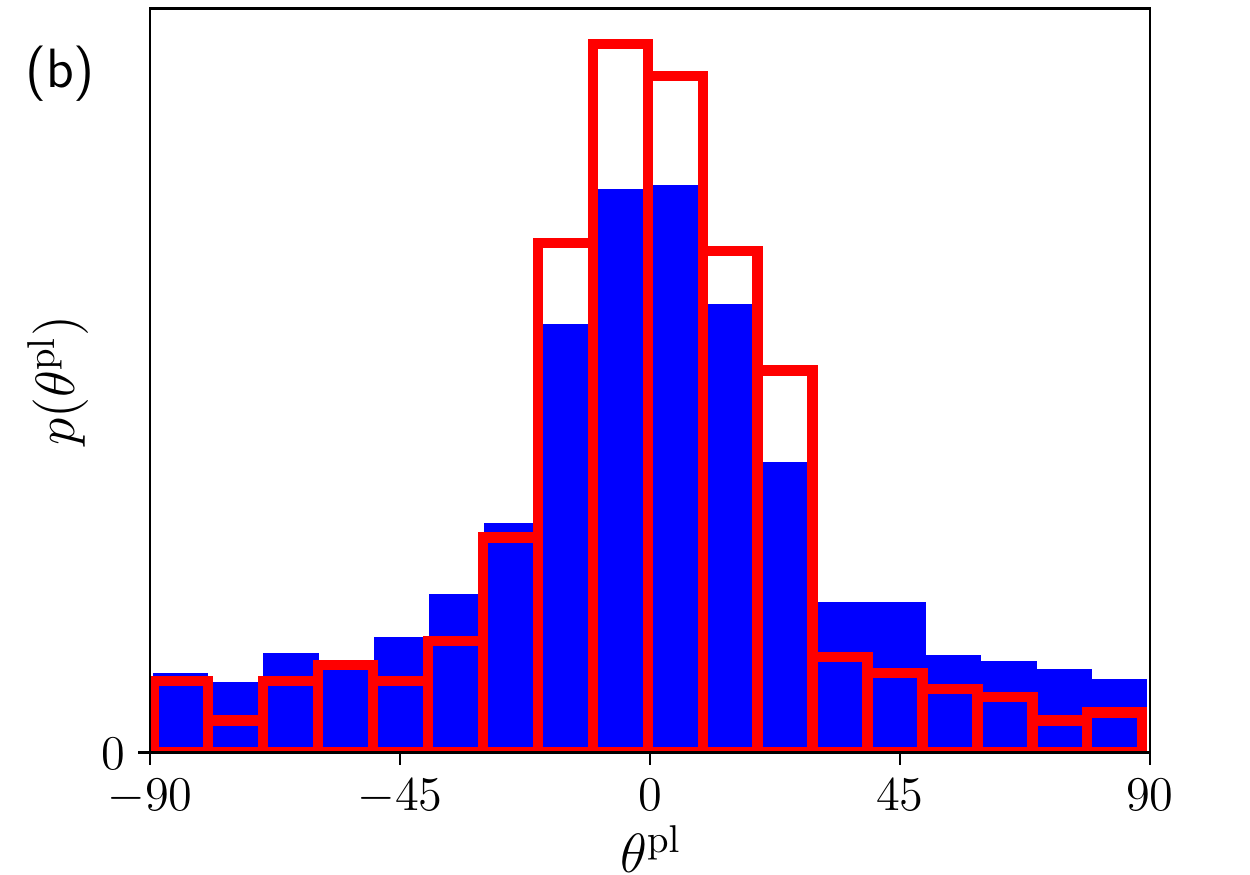} 
\par\end{centering}
\caption{\label{fig:dist_angles_vs_gdot}Distributions of angles of failure $\theta^\mathrm{pl}$ obtained with the MD/Esh method.
\textbf{(a)} Comparison of $p(\theta^\mathrm{pl})$  between distinct shear rates $\dot\gamma$. 
The dashed line represents a normal distribution with standard deviation $\delta \theta^\mathrm{pl}=23^\circ$.
\textbf{(b)} Distribution $p(\theta^\mathrm{pl})$ at $\dot\gamma=10^{-5}$ before (filled blue) and after (red) removing the STs which substantially deviate from their Eshelby fits ($\chi^2>0.5$).}
\end{figure}

\subsubsection{Azimuthal modes of the displacements induced by the STs \label{sec:MDazi}}

A variant of this method may save us the cost of the fitting step. As mentioned in the introduction, the strain field 
$\boldsymbol{\delta \epsilon}$ induced by
the shear part ($\epsilon^\star$) of an ST has a four-fold azimuthal symmetry. Therefore, focusing on $\delta \epsilon_{xy}$ for instance, the $m=4$ azimuthal mode of $\delta \epsilon_{xy}(r)$ contains all information pertaining to the ST orientation (whereas the $m=2$ component results
from the dilational part $\epsilon_v$). In practice, using the auxiliary simulations described above, we compute the local strain around each 
particle (i.e.  the tensor $\boldsymbol{\delta \epsilon}_i$ which minimizes the local non-affine deviations $d^2_\mathrm{min}$ introduced in Sec.~\ref{sec:detection_rearr}), coarse-grain the $xy$-shear strain field
 into boxes of linear size $r_c=3$ (see Fig.~\ref{fig:snapshots}), and compute the azimuthal Fourier modes $c_m$ of the resulting coarse-grained field $\delta \epsilon_{xy}^c$ along a circle of radius $r$ (much larger than the cluster size), viz.,
\begin{equation}
c_m = \int_0^{2\pi} e^{-\mathbf{i} m \theta} \delta \epsilon_{xy}^c(r,\theta) d\theta.
\end{equation} 
Calculating $c_4$ for the quadrupolar strain field and writing it as $c_4=|c_4| e^{\mathbf{i} \phi_4}$,
we find that the angle of failure is related to $\phi_4$ via $\theta^\mathrm{pl} = \phi_4 / 2$.
We call this method Esh/azi.

\subsubsection{Methods exclusively based on the forces or displacements of rearranging particles \label{sec:local_methods}}
The two methods described above involve minimization steps and/or additional (auxiliary) simulations and are therefore numerically costly. To bypass this cost, we will try
to get information on the ST by using only the observed displacements $\delta u_i$ of the particles $i$ within the rearranging cluster. 
A first idea is to compute the internal part $\boldsymbol{\sigma}$ of the local stress tensor: $\boldsymbol{\sigma}= -V^{-1} \sum_i f_i \otimes r_i $, where 
$V$ is the cluster size, the sum runs over all particles $i$ in the cluster, each subjected to an average force $f_i$ and undergoing a displacement $\delta u_i$ between $t$ and $t+\delta t$. The analogue for the
displacements is the tensor $\boldsymbol{\mathcal{M}}= -V^{-1} \sum_i \delta u_i \otimes r_i$. Positions $r_i$ are expressed relative to the
cluster centers of gravity, and the mean force (or displacement) among the ST particles is drawn off the $f_i$ (or $u_i$). 
A yield
angle $\theta^\mathrm{pl}$ can be extracted from these tensors by symmetrising them and writing their deviatoric (traceless) part $\boldsymbol{s}^\mathrm{dev}$ as
\begin{equation}
\boldsymbol{s}^\mathrm{dev} = - \alpha 
\left(\begin{array}{cc}
\sin 2\theta^{\mathrm{pl}} & \cos 2\theta^{\mathrm{pl}} \\
 \cos 2\theta^{\mathrm{pl}}  & -\sin 2\theta^{\mathrm{pl}}
\end{array}\right),
\end{equation}
with a coefficient $\alpha>0$ (the minus sign comes from the sign convention used to define the Cauchy stress). These methods will be referred to as Loc. We have checked that they yield the same result as the inspection of the azimuthal mode $c_4$ of the response of an isotropic homogeneous elastic \emph{continuum}
to the set of pointwise forces $F_i=f_i$, or $F_i\propto \delta u_i$ for the displacement-based version, as computed by means of the Oseen-Burgers tensor. (We have underlined the word \emph{continuum} to insist
on the difference with the MD/azi method).

\section{Characteristics of STs}

In this Section, we employ the method based on fitting rearranging clusters to Eshelby inclusions in order to unveil key characteristics of the rearrangements.
Although STs are often idealized as pure shear transformations, the volumetric deformations are found not to be negligible in practice.
In Fig.~\ref{fig:dist_strain_strenghts}, we report the distributions of the dilational strengths $\pi a^2 \epsilon_v$ and the shear strengths $\pi a^2 \epsilon^\star$
of the STs detected at $\dot\gamma=10^{-5}$,
where $\pi a^2$ is the surface of the inclusion and $\epsilon_v$ and $\epsilon^\star$ were defined in Eq.~\eqref{eq:eigenstrain_Esh}. The corresponding
plots at $\dot\gamma=10^{-4},\,10^{-3}$ are very similar.
As in ref.~\cite{albaret2016mapping}, we observe an exponential distribution of shear strengths, with a typical value around 0.3 here. One should however note that, since the present simulations are not quasi-static, the detected rearrangements (computed over  $\delta t =2$) often do not cover
the whole transformation, which lasts for several time units.

Moving on to the ST orientations, we plot the distribution $p(\theta^\mathrm{pl})$ of angles of failure obtained at the three shear rates in Fig.~\ref{fig:dist_angles_vs_gdot}(a).
We observe no significant sensitivity to the shear rate. Besides, the central part of $p(\theta^\mathrm{pl})$ can be approximated by a normal distribution with standard deviation $\delta \theta^\mathrm{pl}=23^\circ$, but $p(\theta^\mathrm{pl})$ has heavier tails. If we discard the STs for which the elastic response significantly deviates from the Eshelby fit (Fig.~\ref{fig:dist_angles_vs_gdot}(b)), the peak of $p(\theta^\mathrm{pl})$ sharpens
slightly, but this does not strongly affect its shape.

It is interesting to compare these results with those predicted by a mainstream tensorial elasto-plastic model in simple shear \cite{Nicolas2014these}.
The latter also showed a Gaussian-like distribution $p(\theta^\mathrm{pl})$ which was virtually insensitive to the shear rate, but which
was by far narrower than the present ones, with standard deviations of $3-4^\circ$ that could increase up to $\approx 7^\circ$ if
cooperativity in the flow was enhanced by increasing the duration of plastic events or if elasto-plastic blocks were advected along the flow, instead of being static (see Chap.~9.2, p.~111, of \cite{Nicolas2014these}). In these models, angular deviations from the macroscopic shear direction
$\theta^\mathrm{pl}=0$ are exclusively due to cooperative effects, whereby the stress redistributed during an ST (Eq.~\eqref{eq:elastic_prop}) may load other blocks along a direction $\theta^\mathrm{pl}\neq0$, depending on their relative positions. The much broader distribution 
$p(\theta^\mathrm{pl})$
measured in the present atomistic simulations hints at the impact of the granularity of the local medium, which may favor failure along
a direction distinct from that of the local loading.

\section{Comparison between distinct methods to measure ST orientations \label{sec:comparison_methods}}

Having characterized the strengths and orientations of STs, we now discuss to what extent the ST characteristics can be extracted
from methods that do not rely on fits to Eshelby inclusions.

\subsection{Azimuthal mode of the induced strain}
\begin{figure}[t]
\noindent \begin{centering}
\includegraphics[width=\columnwidth]{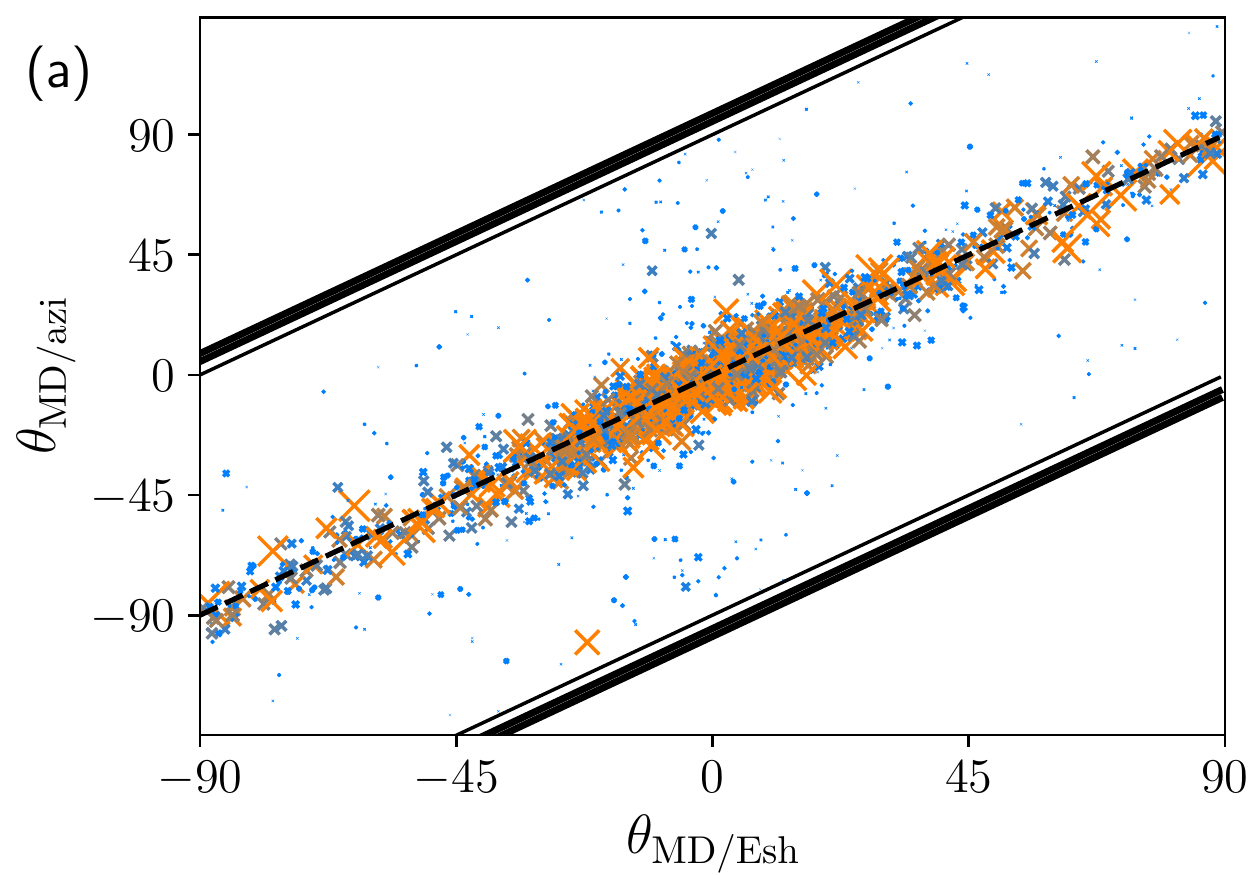} 
\includegraphics[width=\columnwidth]{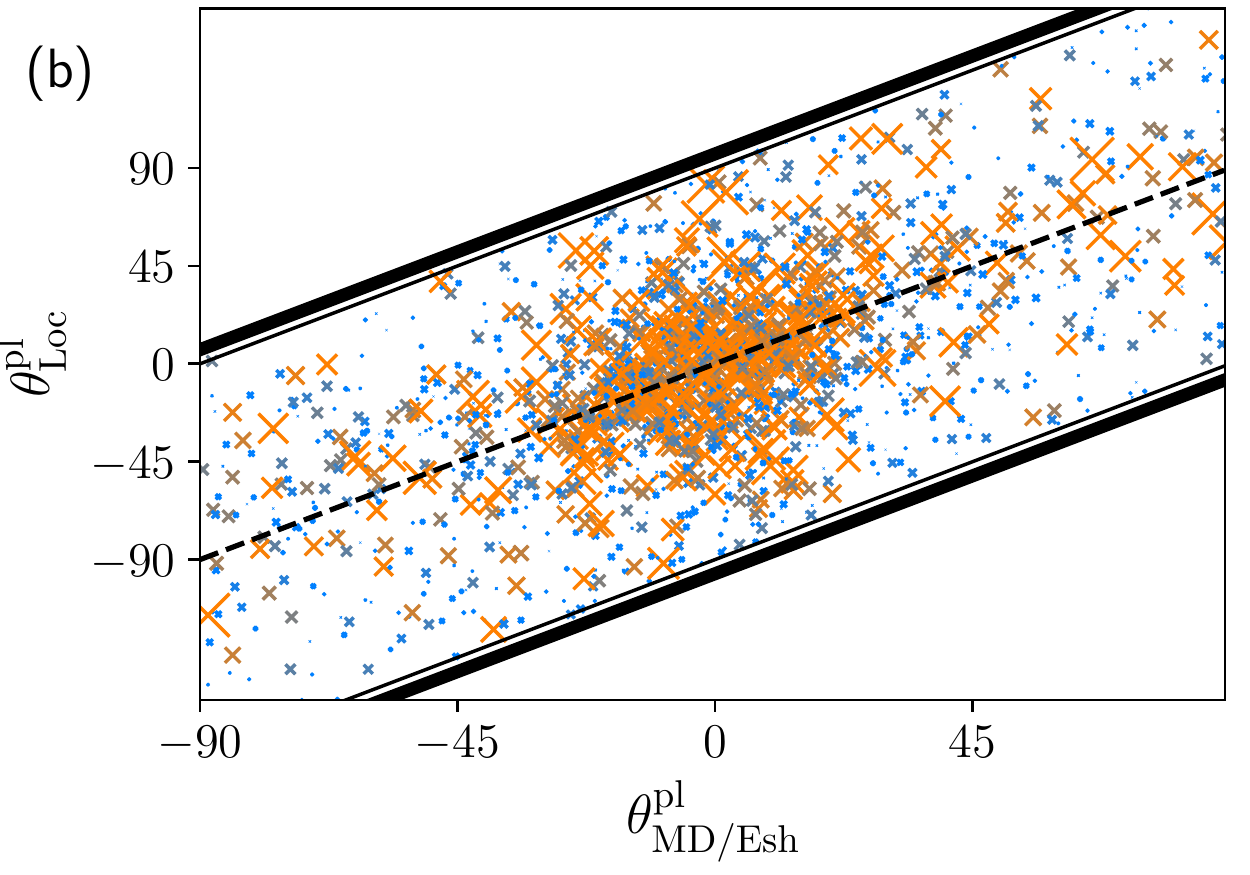} 
\par\end{centering}
\caption{\label{fig:scatters} Scatter plot of angles of failure $\theta^\mathrm{pl}$ measured at $\dot\gamma=10^{-5}$ with \textbf{(a)} the MD/Esh method \emph{vs.} the MD/azi method and \textbf{(b)} the MD/Esh method \emph{vs.} the displacement-based Loc method. Large (orange) crosses refer to STs with good Eshelby fits, while small (blue) crosses indicate poor fits; more precisely, the sizes of the crosses are inversely proportional to the $\chi^2$-deviation from the fit.}
\end{figure}

We start by considering the MD/azi method introduced in Sec.~\ref{sec:MDazi}, which extracts the quadrupolar azimuthal mode of the $xy$-strain (from the
auxiliary MD simulations) on a circle of radius $r$ to determine $\theta^\mathrm{pl}$. The angles of failure $\theta^\mathrm{pl}$ measured 
at distinct $r$ ($r=17$ and $r=23$) are typically within $\pm 10^\circ$ of one another (\emph{data not shown}); there are outliers, but these
very generally correspond to STs that strongly deviate from the Eshelby fits. Hereafter, we fix the radius at $r=17$.
Figure~\ref{fig:scatters}(a) shows that the individual MD/azi angles of failure agree relatively well with those determined with the MD/Esh method used so far,
with absolute differences smaller than $20^\circ$ for STs with reasonable Eshelby fits.

\subsection{Methods based on local forces or displacements}
\begin{figure}[t]
\noindent \begin{centering}
\includegraphics[width=\columnwidth]{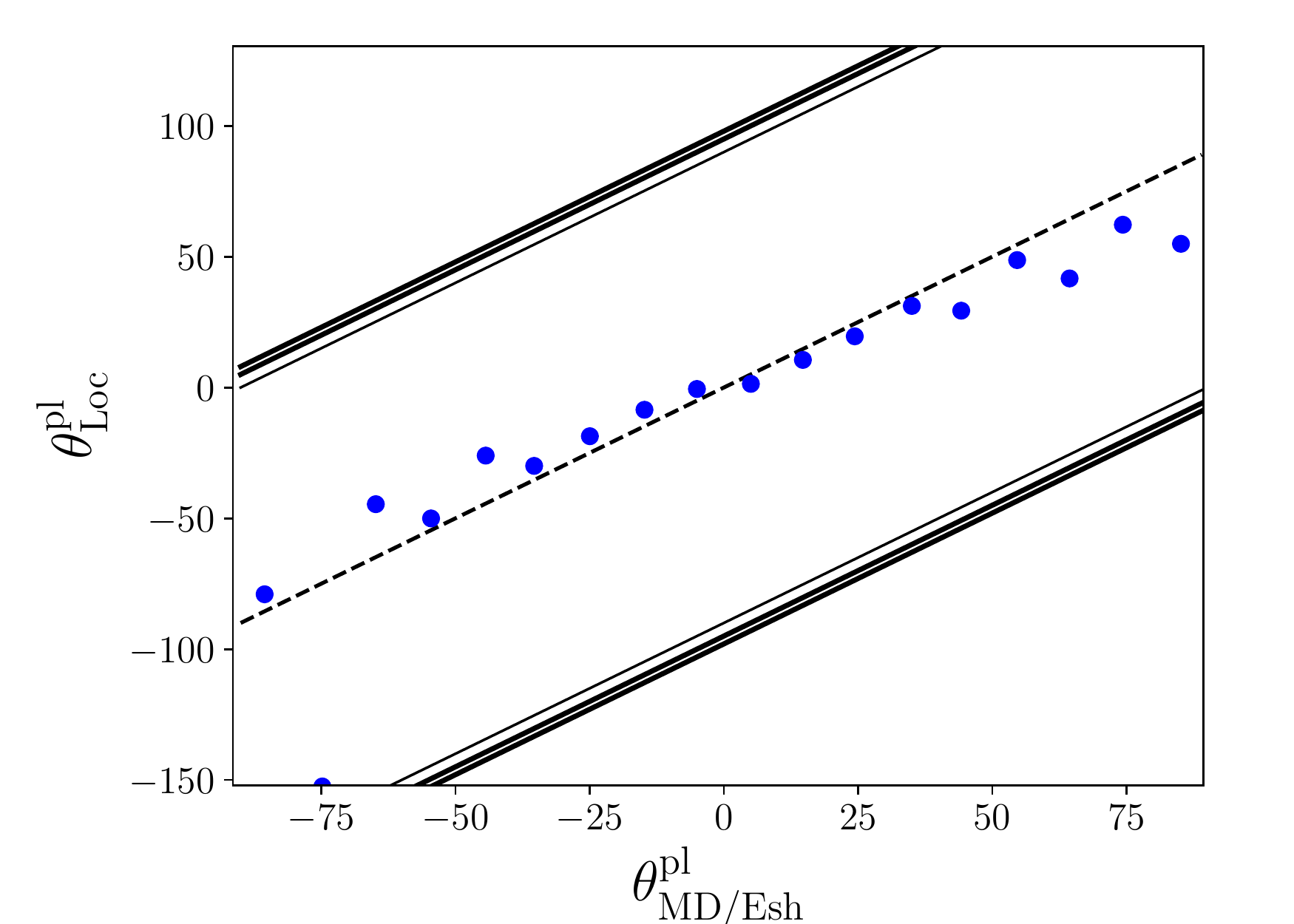}
\par\end{centering}

\caption{\label{fig:binned_angles} Comparison between the angles of failure $\theta^\mathrm{pl}_\mathrm{MD/Esh}$ and $\theta^\mathrm{pl}_\mathrm{Loc}$ measured with the MD/Esh method and the displacement-based local method, respectively.
The STs have been binned into $10^\circ$-wide angular windows, according to the value 
of $\theta^\mathrm{pl}_\mathrm{MD/Esh}$. }
\end{figure}

Turning to the results obtained with local methods (Sec.~\ref{sec:local_methods}),
we report that we have not found any correlation between the MD/Esh angles of failure and those determined with force-based local methods, whether it be the total force $f_i$ or the 'linearized' forces $f_i^{(\mathcal{H})}$ (both being averaged over $\delta t$).
On the other hand, displacement-based local methods broadly agree with MD/Esh, even though this does not immediately transpire from
the scatter plot of Fig.~\ref{fig:scatters}(b). To prove the overall consistency
of the methods despite this large noise, we split the detected STs into $10^\circ$-wide bins depending on their orientation $\theta^\mathrm{pl}_\mathrm{MD/Esh}$ and, for each bin, plot the average angle $\theta^\mathrm{pl}_\mathrm{Loc}$ (measured with the displacement-based local method)
in Fig.~\ref{fig:binned_angles}. On a technical note, one should mention that, to average over angles $\theta_1,...,\theta_n$, we computed the circular average $\mathrm{arg}\left(\sum_j e^{{\bf i} \theta_j} \right)$. 
With these averaged data, the two methods are found to be in good accordance \footnote{Incidentally, note that this is much less the case if STs are binned according to $\theta^\mathrm{pl}_\mathrm{Loc}$.}.

\begin{figure}
\noindent \begin{centering}
\includegraphics[width=\columnwidth]{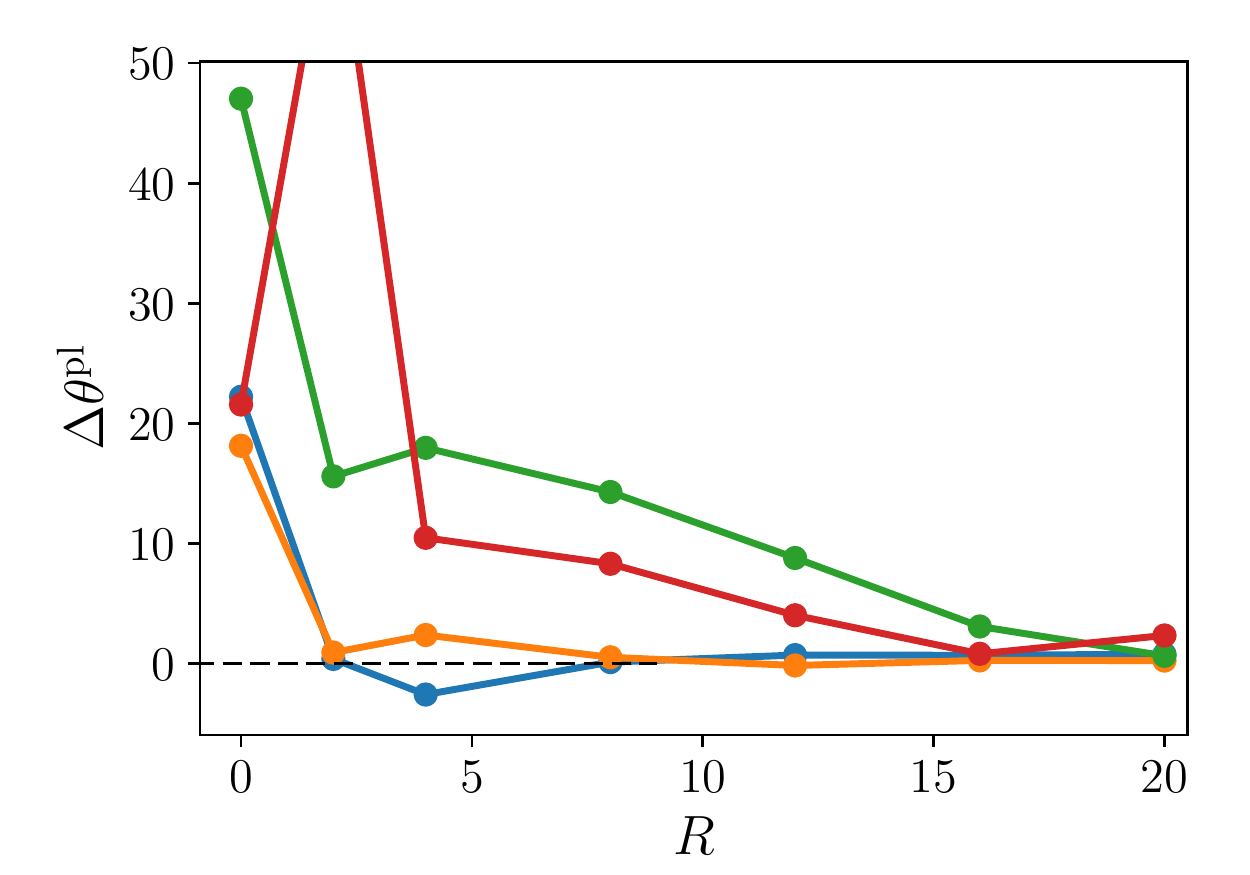}
\par\end{centering}

\caption{\label{fig:incr_radii}Differences $\Delta \theta^\mathrm{pl}$ between the angles of failure found with the local method based on the displacements of all
particles within a distance $R$ of the ST center of gravity (in the auxiliary simulation) and the MD/Esh method for four STs that displayed good Eshelby fits ($\chi^2<1$) but large discrepancies with the Loc method.  For $R=0$, the local method makes only use of the rearranging particles as identified by the kinetic energy threshold. }
\end{figure}

To shed light on the discrepancies in the one-to-one comparison, we extend the local method by including the displacements (measured in the auxiliary
simulation) of all particles within a distance $R$ of the center of gravity of the ST, instead of only the rearranging particles, with the
expectation that both methods converge when $R\to \infty$. In  Fig.~\ref{fig:incr_radii}, we apply this method to STs detected at $\dot\gamma=10^{-5}$
for which a mismatch between $\theta^\mathrm{pl}_\mathrm{Loc}$ and $\theta^\mathrm{pl}_\mathrm{MD/Esh}$ was observed, despite fairly good fits to Eshelby inclusions.
The figure suggests a reasonably quick convergence between the two methods, although the radii $R$ at which
convergence is reached strongly depend on the ST. This implies that the deficiency of
the pristine Loc method stems from its biased selection of too few particles for the computation of the local tensor.

\section{Conclusion}
This paper has introduced and compared three approaches to extract the size and orientation of STs in sheared amorphous solids. Rearranging particles were grouped into clusters based on a threshold criterion for the kinetic energy, which is reliable for athermal 
solids, and their displacements over a small time interval were recorded. Once these clusters are extracted, auxiliary simulations are performed in which the particles taking part in a given ST are displaced and the remainder is relaxed via energy minimization. In the first approach, which we consider to be the most general one, the resulting displacment field is then analyzed by fitting to the ideal "Eshelby" solution for the far-field displacements. In the second method, this fitting is avoided by instead computing the azimutal mode of the (coarse grained) strain field resulting from the ST. Angles of failure obtained from these two methods agree well with each other as long as the Eshelby fit itself is reasonable.

A third and purely local method that avoids auxiliary simulations altogether consists in computing the deviatoric part of the displacement (inertia) tensor after the rearranging clusters have been identified. These angles of failure agree less well with those from Eshelby fits in a point by point comparison, but can be shown to be overall consistent after the noise is reduced through averaging. The inclusion of a larger number of particles improves the agreement between the methods considerably.  In practice, this extended local method is the most efficient one as long as the STs do not overlap.

It will be interesting to compare the angles of failure of STs to the local configurations \emph{prior to} failure,
in particular the direction of the maximal shear stress
and the directional dependence of the local yield stress, which can be measured by deforming a small region embedded in a purely affinely deforming region \cite{Patinet2016,barbot2018local}. Moreover, our results suggest that mesoscopic elastoplastic models  \cite{nicolas2017deformation} should be refined to better describe the deviations from the idealized Eshelby picture observed at the particle scale, and the sensitivity of their predictions to such microscopic details should be examined.

\section*{Acknowledgements}
We thank Jean-Louis Barrat for discussions related to this study. JR is being supported by the Discovery Grant Program of the Natural Sciences and Engineering Research Council of Canada. This research was supported in part by the National Science Foundation under Grant No. NSF PHY11-25915.


\begin{thebibliography}{0}%
\makeatletter
\providecommand \@ifxundefined [1]{%
 \@ifx{#1\undefined}
}%
\providecommand \@ifnum [1]{%
 \ifnum #1\expandafter \@firstoftwo
 \else \expandafter \@secondoftwo
 \fi
}%
\providecommand \@ifx [1]{%
 \ifx #1\expandafter \@firstoftwo
 \else \expandafter \@secondoftwo
 \fi
}%
\providecommand \natexlab [1]{#1}%
\providecommand \enquote  [1]{``#1''}%
\providecommand \bibnamefont  [1]{#1}%
\providecommand \bibfnamefont [1]{#1}%
\providecommand \citenamefont [1]{#1}%
\providecommand \href@noop [0]{\@secondoftwo}%
\providecommand \href [0]{\begingroup \@sanitize@url \@href}%
\providecommand \@href[1]{\@@startlink{#1}\@@href}%
\providecommand \@@href[1]{\endgroup#1\@@endlink}%
\providecommand \@sanitize@url [0]{\catcode `\\12\catcode `\$12\catcode
  `\&12\catcode `\#12\catcode `\^12\catcode `\_12\catcode `\%12\relax}%
\providecommand \@@startlink[1]{}%
\providecommand \@@endlink[0]{}%
\providecommand \url  [0]{\begingroup\@sanitize@url \@url }%
\providecommand \@url [1]{\endgroup\@href {#1}{\urlprefix }}%
\providecommand \urlprefix  [0]{URL }%
\providecommand \Eprint [0]{\href }%
\providecommand \doibase [0]{http://dx.doi.org/}%
\providecommand \selectlanguage [0]{\@gobble}%
\providecommand \bibinfo  [0]{\@secondoftwo}%
\providecommand \bibfield  [0]{\@secondoftwo}%
\providecommand \translation [1]{[#1]}%
\providecommand \BibitemOpen [0]{}%
\providecommand \bibitemStop [0]{}%
\providecommand \bibitemNoStop [0]{.\EOS\space}%
\providecommand \EOS [0]{\spacefactor3000\relax}%
\providecommand \BibitemShut  [1]{\csname bibitem#1\endcsname}%
\let\auto@bib@innerbib\@empty
\end{thebibliography}%


\begin{thebibliography}{31}%
\makeatletter
\providecommand \@ifxundefined [1]{%
 \@ifx{#1\undefined}
}%
\providecommand \@ifnum [1]{%
 \ifnum #1\expandafter \@firstoftwo
 \else \expandafter \@secondoftwo
 \fi
}%
\providecommand \@ifx [1]{%
 \ifx #1\expandafter \@firstoftwo
 \else \expandafter \@secondoftwo
 \fi
}%
\providecommand \natexlab [1]{#1}%
\providecommand \enquote  [1]{``#1''}%
\providecommand \bibnamefont  [1]{#1}%
\providecommand \bibfnamefont [1]{#1}%
\providecommand \citenamefont [1]{#1}%
\providecommand \href@noop [0]{\@secondoftwo}%
\providecommand \href [0]{\begingroup \@sanitize@url \@href}%
\providecommand \@href[1]{\@@startlink{#1}\@@href}%
\providecommand \@@href[1]{\endgroup#1\@@endlink}%
\providecommand \@sanitize@url [0]{\catcode `\\12\catcode `\$12\catcode
  `\&12\catcode `\#12\catcode `\^12\catcode `\_12\catcode `\%12\relax}%
\providecommand \@@startlink[1]{}%
\providecommand \@@endlink[0]{}%
\providecommand \url  [0]{\begingroup\@sanitize@url \@url }%
\providecommand \@url [1]{\endgroup\@href {#1}{\urlprefix }}%
\providecommand \urlprefix  [0]{URL }%
\providecommand \Eprint [0]{\href }%
\providecommand \doibase [0]{http://dx.doi.org/}%
\providecommand \selectlanguage [0]{\@gobble}%
\providecommand \bibinfo  [0]{\@secondoftwo}%
\providecommand \bibfield  [0]{\@secondoftwo}%
\providecommand \translation [1]{[#1]}%
\providecommand \BibitemOpen [0]{}%
\providecommand \bibitemStop [0]{}%
\providecommand \bibitemNoStop [0]{.\EOS\space}%
\providecommand \EOS [0]{\spacefactor3000\relax}%
\providecommand \BibitemShut  [1]{\csname bibitem#1\endcsname}%
\let\auto@bib@innerbib\@empty
\bibitem [{\citenamefont {Argon}\ and\ \citenamefont {Kuo}(1979)}]{Argon1979}%
  \BibitemOpen
  \bibfield  {author} {\bibinfo {author} {\bibfnamefont {A.}~\bibnamefont
  {Argon}}\ and\ \bibinfo {author} {\bibfnamefont {H.}~\bibnamefont {Kuo}},\
  }\href@noop {} {\bibfield  {journal} {\bibinfo  {journal} {Materials Science
  and Engineering}\ }\textbf {\bibinfo {volume} {39}},\ \bibinfo {pages} {101}
  (\bibinfo {year} {1979})}\BibitemShut {NoStop}%
\bibitem [{\citenamefont {Falk}\ and\ \citenamefont {Langer}(1998)}]{Falk1998}%
  \BibitemOpen
  \bibfield  {author} {\bibinfo {author} {\bibfnamefont {M.~L.}\ \bibnamefont
  {Falk}}\ and\ \bibinfo {author} {\bibfnamefont {J.~S.}\ \bibnamefont
  {Langer}},\ }\href@noop {} {\bibfield  {journal} {\bibinfo  {journal}
  {Physical Review E}\ }\textbf {\bibinfo {volume} {57}},\ \bibinfo {pages}
  {7192} (\bibinfo {year} {1998})}\BibitemShut {NoStop}%
\bibitem [{\citenamefont {Maloney}\ and\ \citenamefont
  {Lema{\^\i}tre}(2006)}]{Maloney2006Amorphous}%
  \BibitemOpen
  \bibfield  {author} {\bibinfo {author} {\bibfnamefont {C.~E.}~\bibnamefont
  {Maloney}}\ and\ \bibinfo {author} {\bibfnamefont {A.}~\bibnamefont
  {Lema{\^\i}tre}},\ }\href@noop {} {\bibfield  {journal} {\bibinfo  {journal}
  {Physical Review E}\ }\textbf {\bibinfo {volume} {74}},\ \bibinfo {pages}
  {016118} (\bibinfo {year} {2006})}\BibitemShut {NoStop}%
\bibitem [{\citenamefont {Desmond}\ and\ \citenamefont
  {Weeks}(2015)}]{desmond2015measurement}%
  \BibitemOpen
  \bibfield  {author} {\bibinfo {author} {\bibfnamefont {K.~W.}\ \bibnamefont
  {Desmond}}\ and\ \bibinfo {author} {\bibfnamefont {E.~R.}\ \bibnamefont
  {Weeks}},\ }\href@noop {} {\bibfield  {journal} {\bibinfo  {journal}
  {Physical review letters}\ }\textbf {\bibinfo {volume} {115}},\ \bibinfo
  {pages} {098302} (\bibinfo {year} {2015})}\BibitemShut {NoStop}%
\bibitem [{\citenamefont {Baret}\ \emph {et~al.}(2002)\citenamefont {Baret},
  \citenamefont {Vandembroucq},\ and\ \citenamefont {Roux}}]{Baret2002}%
  \BibitemOpen
  \bibfield  {author} {\bibinfo {author} {\bibfnamefont {J.-C.}\ \bibnamefont
  {Baret}}, \bibinfo {author} {\bibfnamefont {D.}~\bibnamefont {Vandembroucq}},
  \ and\ \bibinfo {author} {\bibfnamefont {S.}~\bibnamefont {Roux}},\
  }\href@noop {} {\bibfield  {journal} {\bibinfo  {journal} {Physical Review
  Letters}\ }\textbf {\bibinfo {volume} {89}},\ \bibinfo {pages} {195506}
  (\bibinfo {year} {2002})}\BibitemShut {NoStop}%
\bibitem [{\citenamefont {Antonaglia}\ \emph {et~al.}(2014)\citenamefont
  {Antonaglia}, \citenamefont {Wright}, \citenamefont {Gu}, \citenamefont
  {Byer}, \citenamefont {Hufnagel}, \citenamefont {LeBlanc}, \citenamefont
  {Uhl},\ and\ \citenamefont {Dahmen}}]{Antonaglia2014}%
  \BibitemOpen
  \bibfield  {author} {\bibinfo {author} {\bibfnamefont {J.}~\bibnamefont
  {Antonaglia}}, \bibinfo {author} {\bibfnamefont {W.~J.}\ \bibnamefont
  {Wright}}, \bibinfo {author} {\bibfnamefont {X.}~\bibnamefont {Gu}}, \bibinfo
  {author} {\bibfnamefont {R.~R.}\ \bibnamefont {Byer}}, \bibinfo {author}
  {\bibfnamefont {T.~C.}\ \bibnamefont {Hufnagel}}, \bibinfo {author}
  {\bibfnamefont {M.}~\bibnamefont {LeBlanc}}, \bibinfo {author} {\bibfnamefont
  {J.~T.}\ \bibnamefont {Uhl}}, \ and\ \bibinfo {author} {\bibfnamefont
  {K.~A.}\ \bibnamefont {Dahmen}},\ }\href@noop {} {\bibfield  {journal}
  {\bibinfo  {journal} {Physical Review Letters}\ }\textbf {\bibinfo {volume}
  {112}},\ \bibinfo {pages} {155501} (\bibinfo {year} {2014})}\BibitemShut
  {NoStop}%
\bibitem [{\citenamefont {Nicolas}\ \emph {et~al.}(2017)\citenamefont
  {Nicolas}, \citenamefont {Ferrero}, \citenamefont {Martens},\ and\
  \citenamefont {Barrat}}]{nicolas2017deformation}%
  \BibitemOpen
  \bibfield  {author} {\bibinfo {author} {\bibfnamefont {A.}~\bibnamefont
  {Nicolas}}, \bibinfo {author} {\bibfnamefont {E.~E.}\ \bibnamefont
  {Ferrero}}, \bibinfo {author} {\bibfnamefont {K.}~\bibnamefont {Martens}}, \
  and\ \bibinfo {author} {\bibfnamefont {J.-L.}\ \bibnamefont {Barrat}},\
  }\href@noop {} {\bibfield  {journal} {\bibinfo  {journal} {arXiv preprint
  arXiv:1708.09194}\ } (\bibinfo {year} {2017})}\BibitemShut {NoStop}%
\bibitem [{\citenamefont {Eshelby}(1957)}]{Eshelby1957}%
  \BibitemOpen
  \bibfield  {author} {\bibinfo {author} {\bibfnamefont {J.}~\bibnamefont
  {Eshelby}},\ }\href@noop {} {\bibfield  {journal} {\bibinfo  {journal}
  {Proceedings of the Royal Society A: Mathematical, Physical and Engineering
  Sciences}\ }\textbf {\bibinfo {volume} {241}},\ \bibinfo {pages} {376}
  (\bibinfo {year} {1957})}\BibitemShut {NoStop}%
\bibitem [{\citenamefont {Puosi}\ \emph {et~al.}(2014)\citenamefont {Puosi},
  \citenamefont {Rottler},\ and\ \citenamefont {Barrat}}]{Puosi2014}%
  \BibitemOpen
  \bibfield  {author} {\bibinfo {author} {\bibfnamefont {F.}~\bibnamefont
  {Puosi}}, \bibinfo {author} {\bibfnamefont {J.}~\bibnamefont {Rottler}}, \
  and\ \bibinfo {author} {\bibfnamefont {J.-L.}\ \bibnamefont {Barrat}},\
  }\href@noop {} {\bibfield  {journal} {\bibinfo  {journal} {Physical Review
  E}\ }\textbf {\bibinfo {volume} {89}},\ \bibinfo {pages} {042302} (\bibinfo
  {year} {2014})}\BibitemShut {NoStop}%
\bibitem [{\citenamefont {Nicolas}\ \emph
  {et~al.}(2014{\natexlab{a}})\citenamefont {Nicolas}, \citenamefont {Martens},
  \citenamefont {Bocquet},\ and\ \citenamefont {Barrat}}]{Nicolas2014u}%
  \BibitemOpen
  \bibfield  {author} {\bibinfo {author} {\bibfnamefont {A.}~\bibnamefont
  {Nicolas}}, \bibinfo {author} {\bibfnamefont {K.}~\bibnamefont {Martens}},
  \bibinfo {author} {\bibfnamefont {L.}~\bibnamefont {Bocquet}}, \ and\
  \bibinfo {author} {\bibfnamefont {J.-L.}\ \bibnamefont {Barrat}},\
  }\href@noop {} {\bibfield  {journal} {\bibinfo  {journal} {Soft Matter}\
  }\textbf {\bibinfo {volume} {10}},\ \bibinfo {pages} {4648} (\bibinfo {year}
  {2014}{\natexlab{a}})}\BibitemShut {NoStop}%
\bibitem [{\citenamefont {Sandfeld}\ and\ \citenamefont
  {Zaiser}(2014)}]{sandfeld2014deformation}%
  \BibitemOpen
  \bibfield  {author} {\bibinfo {author} {\bibfnamefont {S.}~\bibnamefont
  {Sandfeld}}\ and\ \bibinfo {author} {\bibfnamefont {M.}~\bibnamefont
  {Zaiser}},\ }\href@noop {} {\bibfield  {journal} {\bibinfo  {journal}
  {Journal of Statistical Mechanics: Theory and Experiment}\ }\textbf {\bibinfo
  {volume} {2014}},\ \bibinfo {pages} {P03014} (\bibinfo {year}
  {2014})}\BibitemShut {NoStop}%
\bibitem [{\citenamefont {Talamali}\ \emph {et~al.}(2011)\citenamefont
  {Talamali}, \citenamefont {Pet\"{a}j\"{a}}, \citenamefont {Vandembroucq},\
  and\ \citenamefont {Roux}}]{Talamali2011}%
  \BibitemOpen
  \bibfield  {author} {\bibinfo {author} {\bibfnamefont {M.}~\bibnamefont
  {Talamali}}, \bibinfo {author} {\bibfnamefont {V.}~\bibnamefont
  {Pet\"{a}j\"{a}}}, \bibinfo {author} {\bibfnamefont {D.}~\bibnamefont
  {Vandembroucq}}, \ and\ \bibinfo {author} {\bibfnamefont {S.}~\bibnamefont
  {Roux}},\ }\href@noop {} {\bibfield  {journal} {\bibinfo  {journal} {Physical
  Review E}\ }\textbf {\bibinfo {volume} {84}},\ \bibinfo {pages} {016115}
  (\bibinfo {year} {2011})}\BibitemShut {NoStop}%
\bibitem [{\citenamefont {Budrikis}\ and\ \citenamefont
  {Zapperi}(2013)}]{Budrikis2013}%
  \BibitemOpen
  \bibfield  {author} {\bibinfo {author} {\bibfnamefont {Z.}~\bibnamefont
  {Budrikis}}\ and\ \bibinfo {author} {\bibfnamefont {S.}~\bibnamefont
  {Zapperi}},\ }\href@noop {} {\bibfield  {journal} {\bibinfo  {journal}
  {Physical Review E}\ }\textbf {\bibinfo {volume} {88}},\ \bibinfo {pages}
  {062403} (\bibinfo {year} {2013})}\BibitemShut {NoStop}%
\bibitem [{\citenamefont {Homer}\ and\ \citenamefont
  {Schuh}(2009)}]{Homer2009}%
  \BibitemOpen
  \bibfield  {author} {\bibinfo {author} {\bibfnamefont {E.}~\bibnamefont
  {Homer}}\ and\ \bibinfo {author} {\bibfnamefont {C.}~\bibnamefont {Schuh}},\
  }\href@noop {} {\bibfield  {journal} {\bibinfo  {journal} {Acta Materialia}\
  }\textbf {\bibinfo {volume} {57}},\ \bibinfo {pages} {2823} (\bibinfo {year}
  {2009})}\BibitemShut {NoStop}%
\bibitem [{\citenamefont {Schall}\ \emph {et~al.}(2007)\citenamefont {Schall},
  \citenamefont {Weitz},\ and\ \citenamefont {Spaepen}}]{Schall2007}%
  \BibitemOpen
  \bibfield  {author} {\bibinfo {author} {\bibfnamefont {P.}~\bibnamefont
  {Schall}}, \bibinfo {author} {\bibfnamefont {D.}~\bibnamefont {Weitz}}, \
  and\ \bibinfo {author} {\bibfnamefont {F.}~\bibnamefont {Spaepen}},\
  }\href@noop {} {\bibfield  {journal} {\bibinfo  {journal} {Science (New York,
  N.Y.)}\ }\textbf {\bibinfo {volume} {318}},\ \bibinfo {pages} {1895}
  (\bibinfo {year} {2007})}\BibitemShut {NoStop}%
\bibitem [{\citenamefont {Ma}\ \emph {et~al.}(2015)\citenamefont {Ma},
  \citenamefont {Ye}, \citenamefont {Peng}, \citenamefont {Wen},\ and\
  \citenamefont {Zhang}}]{ma2015nanoindentation}%
  \BibitemOpen
  \bibfield  {author} {\bibinfo {author} {\bibfnamefont {Y.}~\bibnamefont
  {Ma}}, \bibinfo {author} {\bibfnamefont {J.}~\bibnamefont {Ye}}, \bibinfo
  {author} {\bibfnamefont {G.}~\bibnamefont {Peng}}, \bibinfo {author}
  {\bibfnamefont {D.}~\bibnamefont {Wen}}, \ and\ \bibinfo {author}
  {\bibfnamefont {T.}~\bibnamefont {Zhang}},\ }\href@noop {} {\bibfield
  {journal} {\bibinfo  {journal} {Materials Science and Engineering: A}\
  }\textbf {\bibinfo {volume} {627}},\ \bibinfo {pages} {153} (\bibinfo {year}
  {2015})}\BibitemShut {NoStop}%
\bibitem [{\citenamefont {Albaret}\ \emph {et~al.}(2016)\citenamefont
  {Albaret}, \citenamefont {Tanguy}, \citenamefont {Boioli},\ and\
  \citenamefont {Rodney}}]{albaret2016mapping}%
  \BibitemOpen
  \bibfield  {author} {\bibinfo {author} {\bibfnamefont {T.}~\bibnamefont
  {Albaret}}, \bibinfo {author} {\bibfnamefont {A.}~\bibnamefont {Tanguy}},
  \bibinfo {author} {\bibfnamefont {F.}~\bibnamefont {Boioli}}, \ and\ \bibinfo
  {author} {\bibfnamefont {D.}~\bibnamefont {Rodney}},\ }\href@noop {}
  {\bibfield  {journal} {\bibinfo  {journal} {Physical Review E}\ }\textbf
  {\bibinfo {volume} {93}},\ \bibinfo {pages} {053002} (\bibinfo {year}
  {2016})}\BibitemShut {NoStop}%
\bibitem [{\citenamefont {Nicolas}\ \emph
  {et~al.}(2014{\natexlab{b}})\citenamefont {Nicolas}, \citenamefont
  {Rottler},\ and\ \citenamefont {Barrat}}]{Nicolas2014s}%
  \BibitemOpen
  \bibfield  {author} {\bibinfo {author} {\bibfnamefont {A.}~\bibnamefont
  {Nicolas}}, \bibinfo {author} {\bibfnamefont {J.}~\bibnamefont {Rottler}}, \
  and\ \bibinfo {author} {\bibfnamefont {J.-L.}\ \bibnamefont {Barrat}},\
  }\href@noop {} {\bibfield  {journal} {\bibinfo  {journal} {The European
  Physical Journal E}\ }\textbf {\bibinfo {volume} {37}},\ \bibinfo {eid} {50}
  (\bibinfo {year} {2014}{\natexlab{b}})}\BibitemShut {NoStop}%
\bibitem [{\citenamefont {Lema{\^\i}tre}(2015)}]{lemaitre2015tensorial}%
  \BibitemOpen
  \bibfield  {author} {\bibinfo {author} {\bibfnamefont {A.}~\bibnamefont
  {Lema{\^\i}tre}},\ }\href@noop {} {\bibfield  {journal} {\bibinfo  {journal}
  {The Journal of chemical physics}\ }\textbf {\bibinfo {volume} {143}},\
  \bibinfo {pages} {164515} (\bibinfo {year} {2015})}\BibitemShut {NoStop}%
\bibitem [{\citenamefont {Schwarz}(1965)}]{schwarz1965rearrangements}%
  \BibitemOpen
  \bibfield  {author} {\bibinfo {author} {\bibfnamefont {H.}~\bibnamefont
  {Schwarz}},\ }\href@noop {} {\bibfield  {journal} {\bibinfo  {journal}
  {Recueil des travaux chimiques des Pays-Bas}\ }\textbf {\bibinfo {volume}
  {84}},\ \bibinfo {pages} {771} (\bibinfo {year} {1965})}\BibitemShut
  {NoStop}%
\bibitem [{\citenamefont {Princen}(1983)}]{Princen1983}%
  \BibitemOpen
  \bibfield  {author} {\bibinfo {author} {\bibfnamefont {H.}~\bibnamefont
  {Princen}},\ }\href@noop {} {\bibfield  {journal} {\bibinfo  {journal}
  {Journal of Colloid and interface science}\ }\textbf {\bibinfo {volume}
  {91}},\ \bibinfo {pages} {160} (\bibinfo {year} {1983})}\BibitemShut
  {NoStop}%
\bibitem [{\citenamefont {Biance}\ \emph {et~al.}(2011)\citenamefont {Biance},
  \citenamefont {Calbry-Muzyka}, \citenamefont {H\"ohler},\ and\ \citenamefont
  {Cohen-Addad}}]{biance2011strain}%
  \BibitemOpen
  \bibfield  {author} {\bibinfo {author} {\bibfnamefont {A.-L.}\ \bibnamefont
  {Biance}}, \bibinfo {author} {\bibfnamefont {A.}~\bibnamefont
  {Calbry-Muzyka}}, \bibinfo {author} {\bibfnamefont {R.}~\bibnamefont
  {H\"ohler}}, \ and\ \bibinfo {author} {\bibfnamefont {S.}~\bibnamefont
  {Cohen-Addad}},\ }\href@noop {} {\bibfield  {journal} {\bibinfo  {journal}
  {Langmuir}\ }\textbf {\bibinfo {volume} {28}},\ \bibinfo {pages} {111}
  (\bibinfo {year} {2011})}\BibitemShut {NoStop}%
\bibitem [{\citenamefont {Choi}\ \emph {et~al.}(2012)\citenamefont {Choi},
  \citenamefont {Zhao}, \citenamefont {Kim}, \citenamefont {Yoo}, \citenamefont
  {Suh}, \citenamefont {Ramamurty},\ and\ \citenamefont
  {Jang}}]{choi2012indentation}%
  \BibitemOpen
  \bibfield  {author} {\bibinfo {author} {\bibfnamefont {I.-C.}\ \bibnamefont
  {Choi}}, \bibinfo {author} {\bibfnamefont {Y.}~\bibnamefont {Zhao}}, \bibinfo
  {author} {\bibfnamefont {Y.-J.}\ \bibnamefont {Kim}}, \bibinfo {author}
  {\bibfnamefont {B.-G.}\ \bibnamefont {Yoo}}, \bibinfo {author} {\bibfnamefont
  {J.-Y.}\ \bibnamefont {Suh}}, \bibinfo {author} {\bibfnamefont
  {U.}~\bibnamefont {Ramamurty}}, \ and\ \bibinfo {author} {\bibfnamefont
  {J.-i.}\ \bibnamefont {Jang}},\ }\href@noop {} {\bibfield  {journal}
  {\bibinfo  {journal} {Acta Materialia}\ }\textbf {\bibinfo {volume} {60}},\
  \bibinfo {pages} {6862} (\bibinfo {year} {2012})}\BibitemShut {NoStop}%
\bibitem [{\citenamefont {Boioli}\ \emph {et~al.}(2017)\citenamefont {Boioli},
  \citenamefont {Albaret},\ and\ \citenamefont {Rodney}}]{boioli2017shear}%
  \BibitemOpen
  \bibfield  {author} {\bibinfo {author} {\bibfnamefont {F.}~\bibnamefont
  {Boioli}}, \bibinfo {author} {\bibfnamefont {T.}~\bibnamefont {Albaret}}, \
  and\ \bibinfo {author} {\bibfnamefont {D.}~\bibnamefont {Rodney}},\
  }\href@noop {} {\bibfield  {journal} {\bibinfo  {journal} {Physical Review
  E}\ }\textbf {\bibinfo {volume} {95}},\ \bibinfo {pages} {033005} (\bibinfo
  {year} {2017})}\BibitemShut {NoStop}%
\bibitem [{\citenamefont {Nicolas}\ \emph {et~al.}(2016)\citenamefont
  {Nicolas}, \citenamefont {Barrat},\ and\ \citenamefont
  {Rottler}}]{nicolas2016effects}%
  \BibitemOpen
  \bibfield  {author} {\bibinfo {author} {\bibfnamefont {A.}~\bibnamefont
  {Nicolas}}, \bibinfo {author} {\bibfnamefont {J.-L.}\ \bibnamefont {Barrat}},
  \ and\ \bibinfo {author} {\bibfnamefont {J.}~\bibnamefont {Rottler}},\
  }\href@noop {} {\bibfield  {journal} {\bibinfo  {journal} {Physical Review
  Letters}\ }\textbf {\bibinfo {volume} {116}},\ \bibinfo {pages} {058303}
  (\bibinfo {year} {2016})}\BibitemShut {NoStop}%
\bibitem [{\citenamefont {Chikkadi}\ and\ \citenamefont
  {Schall}(2012)}]{Chikkadi2012b}%
  \BibitemOpen
  \bibfield  {author} {\bibinfo {author} {\bibfnamefont {V.}~\bibnamefont
  {Chikkadi}}\ and\ \bibinfo {author} {\bibfnamefont {P.}~\bibnamefont
  {Schall}},\ }\href@noop {} {\bibfield  {journal} {\bibinfo  {journal}
  {Physical Review E}\ }\textbf {\bibinfo {volume} {85}},\ \bibinfo {pages}
  {031402} (\bibinfo {year} {2012})}\BibitemShut {NoStop}%
\bibitem [{\citenamefont {Jin}\ \emph {et~al.}(2017)\citenamefont {Jin},
  \citenamefont {Zhang}, \citenamefont {Li}, \citenamefont {Xu}, \citenamefont
  {Hu},\ and\ \citenamefont {Keer}}]{jin2017displacement}%
  \BibitemOpen
  \bibfield  {author} {\bibinfo {author} {\bibfnamefont {X.}~\bibnamefont
  {Jin}}, \bibinfo {author} {\bibfnamefont {X.}~\bibnamefont {Zhang}}, \bibinfo
  {author} {\bibfnamefont {P.}~\bibnamefont {Li}}, \bibinfo {author}
  {\bibfnamefont {Z.}~\bibnamefont {Xu}}, \bibinfo {author} {\bibfnamefont
  {Y.}~\bibnamefont {Hu}}, \ and\ \bibinfo {author} {\bibfnamefont {L.~M.}\
  \bibnamefont {Keer}},\ }\href@noop {} {\bibfield  {journal} {\bibinfo
  {journal} {Journal of Applied Mechanics}\ }\textbf {\bibinfo {volume} {84}},\
  \bibinfo {pages} {074501} (\bibinfo {year} {2017})}\BibitemShut {NoStop}%
\bibitem [{\citenamefont {Nicolas}(2014)}]{Nicolas2014these}%
  \BibitemOpen
  \bibfield  {author} {\bibinfo {author} {\bibfnamefont {A.}~\bibnamefont
  {Nicolas}},\ }\emph {\bibinfo {title} {The Flow of Amorphous Solids:
  Elastoplastic Models and Mode-Coupling Approach}},\ \href@noop {} {Ph.D.
  thesis},\ \bibinfo  {school} {Universit\'e de Grenoble} (\bibinfo {year}
  {2014})\BibitemShut {NoStop}%
\bibitem [{Note1()}]{Note1}%
  \BibitemOpen
  \bibinfo {note} {Incidentally, note that this is much less the case if STs
  are binned according to $\theta ^\protect \mathrm {pl}_\protect \mathrm
  {Loc}$.}\BibitemShut {Stop}%
\bibitem [{\citenamefont {Patinet}\ \emph {et~al.}(2016)\citenamefont
  {Patinet}, \citenamefont {Vandembroucq},\ and\ \citenamefont
  {Falk}}]{Patinet2016}%
  \BibitemOpen
  \bibfield  {author} {\bibinfo {author} {\bibfnamefont {S.}~\bibnamefont
  {Patinet}}, \bibinfo {author} {\bibfnamefont {D.}~\bibnamefont
  {Vandembroucq}}, \ and\ \bibinfo {author} {\bibfnamefont {M.~L.}\
  \bibnamefont {Falk}},\ }\href@noop {} {\bibfield  {journal} {\bibinfo
  {journal} {Phys. Rev. Lett}\ }\textbf {\bibinfo {volume} {117}},\ \bibinfo
  {pages} {045501} (\bibinfo {year} {2016})}\BibitemShut {NoStop}%
\bibitem [{\citenamefont {Barbot}\ \emph {et~al.}(2018)\citenamefont {Barbot},
  \citenamefont {Lerbinger}, \citenamefont {Hernandez-Garcia}, \citenamefont
  {Garc{\'\i}a-Garc{\'\i}a}, \citenamefont {Falk}, \citenamefont
  {Vandembroucq},\ and\ \citenamefont {Patinet}}]{barbot2018local}%
  \BibitemOpen
  \bibfield  {author} {\bibinfo {author} {\bibfnamefont {A.}~\bibnamefont
  {Barbot}}, \bibinfo {author} {\bibfnamefont {M.}~\bibnamefont {Lerbinger}},
  \bibinfo {author} {\bibfnamefont {A.}~\bibnamefont {Hernandez-Garcia}},
  \bibinfo {author} {\bibfnamefont {R.}~\bibnamefont
  {Garc{\'\i}a-Garc{\'\i}a}}, \bibinfo {author} {\bibfnamefont {M.~L.}\
  \bibnamefont {Falk}}, \bibinfo {author} {\bibfnamefont {D.}~\bibnamefont
  {Vandembroucq}}, \ and\ \bibinfo {author} {\bibfnamefont {S.}~\bibnamefont
  {Patinet}},\ }\href@noop {} {\bibfield  {journal} {\bibinfo  {journal}
  {Physical Review E}\ }\textbf {\bibinfo {volume} {97}},\ \bibinfo {pages}
  {033001} (\bibinfo {year} {2018})}\BibitemShut {NoStop}%
\end{thebibliography}
\end{document}